\documentclass[fleqn,usenatbib]{mnras}

\usepackage{newtxtext,newtxmath}
\usepackage{supertabular}

\usepackage[T1]{fontenc}
\usepackage{ae,aecompl}

\usepackage{lscape}

\usepackage{graphicx}	
\usepackage{amsmath}	
\usepackage{amssymb}	
\usepackage{pdflscape}
\usepackage{longtable}
\usepackage{natbib}
\usepackage{booktabs}

\title[Galactic census of RRATs]{The RRATalog: a Galactic census of rotating radio transients}

\author[Agarwal et al.]{Devansh Agarwal$^{1,2}$, Evan F. Lewis$^{1,2}$, Duncan R. Lorimer$^{1,2}$\thanks{E-mail: duncan.lorimer@mail.wvu.edu},  Maura A. McLaughlin$^{1,2}$,  Bingyi Cui$^3$, 
\newauthor
Anna Turner$^{1,4}$ and Natasha McMann$^5$\\
$^{1}$West Virginia University, Department of Physics and Astronomy, Morgantown, WV 26506-6315, USA\\
$^{2}$Center for Gravitational Waves and Cosmology, West Virginia University, Chestnut Ridge Research Building, Morgantown, WV 26505-6315, USA\\
$^{3}$Shanghai Science and Technology Museum, 2000 Century Avenue, Pudong New Area, Shanghai, 200127, China\\
$^{4}$Center for KINETIC Plasma Physics, West Virginia University, Morgantown,
West Virginia 26506, USA\\
$^{5}$Labry School of Science, Technology, and Business, Cumberland University, Memorial Hall, 1 Cumberland Square, Lebanon, TN 37087-3408, USA}

\date{Accepted 2026 April 23. Received 2026 April 23; in original form 2026 April 1}

\pubyear{2026}

\newcommand{\psrpoppy}{\textsc{PsrPopPy2}}

\begin{document}
\label{firstpage}
\pagerange{\pageref{firstpage}--\pageref{lastpage}}
\maketitle

\begin{abstract}
Rotating radio transients (RRATs) represent a significant but poorly understood component of the Galactic neutron star population, characterized by sporadic emission first detectable only through single-pulse searches. We present the RRATalog, an up-to-date catalogue of 335 RRATs, and utilize a uniform sample of RRATs discovered in four Parkes telescope surveys to model their Galactic population. Accounting in detail for observational selection effects, we find a radial density profile similar to pulsars, but identify a significantly steeper luminosity function (power-law index $\alpha \simeq -1.3$) than previously assumed. For sources beaming towards Earth, we estimate $34000 \pm 1600$ potentially observable RRATs above a peak luminosity of 30 mJy kpc$^2$. At these high luminosities, the RRAT population is comparable in size to that of canonical pulsars.  Consistent with the observed distribution, the underlying period distribution is significantly shifted toward longer periods compared to canonical pulsars, suggesting RRATs represent a more evolved population. We find evidence for a turnover in the luminosity function below 30 mJy kpc$^2$, and predict that the total number of potentially observable RRATs is $\lesssim 70,000$. Applying the Tauris \& Manchester beaming model, we estimate the total Galactic RRAT population to be  $\lesssim { 400,000}$. The implied birth rate of $\lesssim 1.4$ RRATs per century is consistent with the Galactic core-collapse supernova rate, suggesting RRATs can be reconciled with known progenitor rates without requiring a separate evolutionary origin. We provide predictions for RRAT discoveries in ongoing and future surveys.
\end{abstract}


\section{Introduction}
Rotating radio transients (hereafter RRATs) were discovered during single-pulse searches of Parkes multibeam pulsar survey data \citep{McLaughlin2006}. { Throughout this work, we define} RRATs { as} rotating neutron stars { that were discovered only through their single pulses because their time-averaged emission was too faint to be detectable via   periodicity searches}. { We define} canonical pulsars, { on the other hand, as rotating neutron stars which were discovered by periodicity searches.} { As a result,} RRATs { are} very difficult to both discover and monitor { because, using a system with comparable sensitivity to the discovery instrument, they are only sporadically seen}. This paper conducts a census of the present sample of 335 RRATs and, by modeling their detectability, makes inferences about the underlying RRAT population.

While it is generally accepted { \citep[see, e.g.,][]{2006ApJ...645L.149W,BS10,2011BASI...39..333K}} that RRATs are a manifestation of the pulsar phenomenon, many theories have been put forward to explain why RRATs show different emission behavior from other pulsars. RRATs may be just one extreme of the neutron star intermittency spectrum, which sits as the extension of nulling pulsars with extremely high nulling fractions \citep{BS13}. \citet{Li2006} suggested that such intermittency is caused by material fallback from a supernova debris disk. Another mechanism is the radio emission from infalling circumstellar material affecting the charge density in the magnetosphere \citep{CS08}. It has also been suggested that some RRAT emission could be produced through similar mechanisms to fast radio bursts \citep[see, e.g.,][]{2016arXiv160806952R}.

 \citet{Nipuni2011} characterized the timing periodicities and pulse clustering and found periodicity in some RRAT burst rates on timescales ranging between $\sim$0.5~hrs and $\sim$5~yrs. They also found burst times to be consistent with  a random distribution. \citet{2017ApJ...840....5C} present the timing solutions for eight RRATs and show log-normal distribution of the pulse amplitudes with some RRATs showing additional power-law tails. \citet{Brent2018} studied the spectral index and wait time distributions of three RRATs and found single-pulse spectral indices ranging from --7 to +4 and some evidence for pulse clustering i.e., instances of two, three or more consecutive pulses. { This range of spectral indices is larger than that found for PSRs B0329+54 and B1133+16 \citep{2003A&A...407..655K}, but  smaller than that seen for the Crab pulsar \citep{2010A&A...515A..36K}.}
 \citet{Mitch2018} found that  the single-pulse amplitude distributions of RRATs and pulsars were quite similar, suggesting a common emission mechanism. 
Recent high-sensitivity observations have further blurred the line between these populations. Notably, 
the Five Hundred Metre Aperture Spherical Telescope Galactic Plane Pulsar Snapshot (FAST GPPS) survey \citep{2023RAA....23j4001Z} demonstrated that many objects previously classified as RRATs appear as weak canonical pulsars when observed with greater sensitivity. As pointed out earlier by \citet{2006ApJ...645L.149W}, this suggests that a significant fraction of the RRAT population may be the high-amplitude tail of a standard pulsar emission distribution, rather than a physically distinct class of intermittent rotators.

\begin{figure*}
    \centering
    \includegraphics[width=0.9\textwidth]{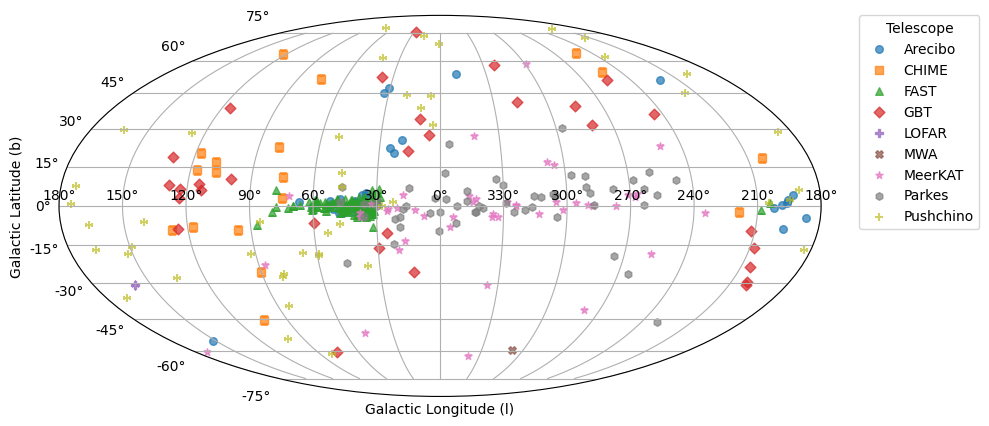}
    \caption{Mollweide projection showing the Galactic distribution of RRATs. The symbols in the legend represent the discovery telescope.}
    \label{fig:sky_distribution}
\end{figure*}

\begin{figure*}
    \centering
    \includegraphics[width=\textwidth]{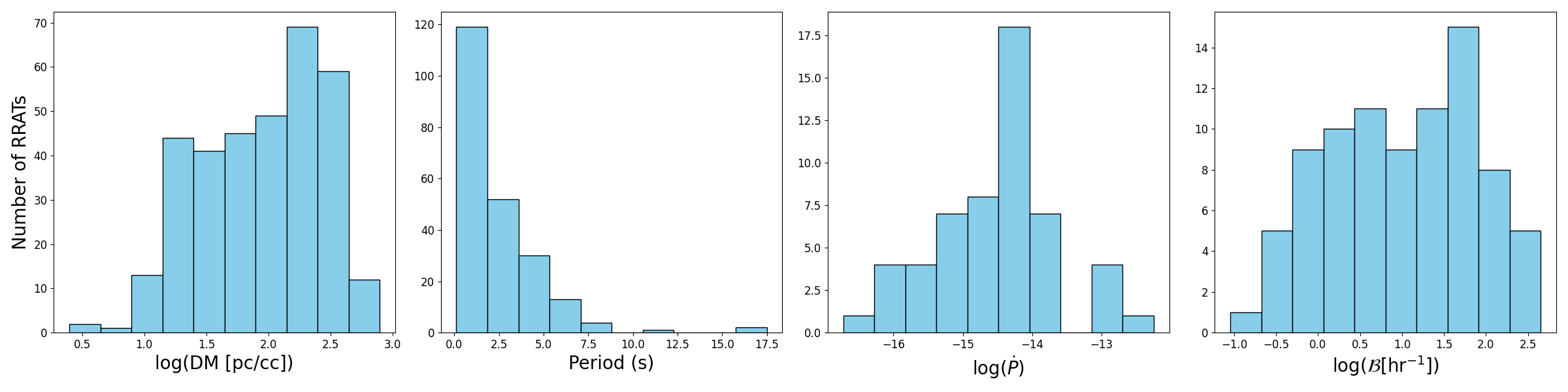}
    \caption{Histograms showing the distributions of observed quantities for RRATs as a function of dispersion measure (DM), spin period, period derivative ($\dot{P}$) and burst rate (${\cal B})$. These distributions have not been corrected for observational selection.}
    \label{fig:observed_hist}
\end{figure*}

The sheer number of RRATs also poses a significant challenge to our understanding of neutron star evolution. If RRATs represent a distinct, long-lived population of objects, their estimated birth rate may be as high as 2 per century, potentially rivaling or even exceeding the Galactic supernova rate of 2--3 per century \citep{2021NewA...8301498R,2025A&A...698A.306M} when combined with canonical pulsars \citep{2008MNRAS.391.2009K}. This `birth rate problem' suggests that many RRATs may instead be a transient evolutionary phase of other neutron star populations, such as high-magnetic-field pulsars, canonical pulsars or magnetars. Accurately modeling the Galactic population of RRATs is therefore essential not only for survey predictions but also for reconciling these objects with the known supernova rate. { One of the main aims of this paper is to revisit the RRAT birth rate and assess its contribution to the neutron star population.}

\citet{Lorimer2006}, hereafter LFL06, conducted a detailed analysis of the  population of canonical pulsars using Monte Carlo simulations. Beyond modeling the inverse-square law, LFL06 carefully constructed survey models to take into account selection effects that are a result of instrumental limitations in the observing system and detection limits caused by propagation through the interstellar medium. Their simulations produce a model of the pulsar population such that, when passed through the survey models, the resultant detected population closely mimics the observed pulsar population.

We build upon the methods described in LFL06 to construct a model of the Galactic population of RRATs. Preliminary results from this work were used and briefly discussed in the context of an Arecibo survey \citep{Patel2018} and the Australian Square Kilometre Array Pathfinder \citep{Qiu_2019} surveys. 
In addition to predicting yields of future surveys and { to} better understand neutron star populations and emission mechanisms, a better understanding of the RRAT population is essential to further investigate the question of whether some of the RRATs with very high dispersion measures are in fact fast radio bursts \citep[for a discussion, see, e.g.,][]{2016MNRAS.459.1360K}.

The rest of this paper is organised as follows. We introduce the catalogue of RRATs in \S\ref{sec:2} and the revised pulsar population package \psrpoppy\,to model the RRAT population in \S\ref{sec:3}. We detail the Monte Carlo methods to generate the Galactic population of RRATs in \S\ref{sec:4}. We use our model to give predictions for upcoming surveys in \S\ref{sec:5} and provide a further discussion of our simulations in \S\ref{sec:6}. We summarise and present suggestions for future work in \S\ref{sec:7}.

\section{The \textsc{RRATalog}}
\label{sec:2}

\begin{figure}
    \centering
    \includegraphics[width=\columnwidth]{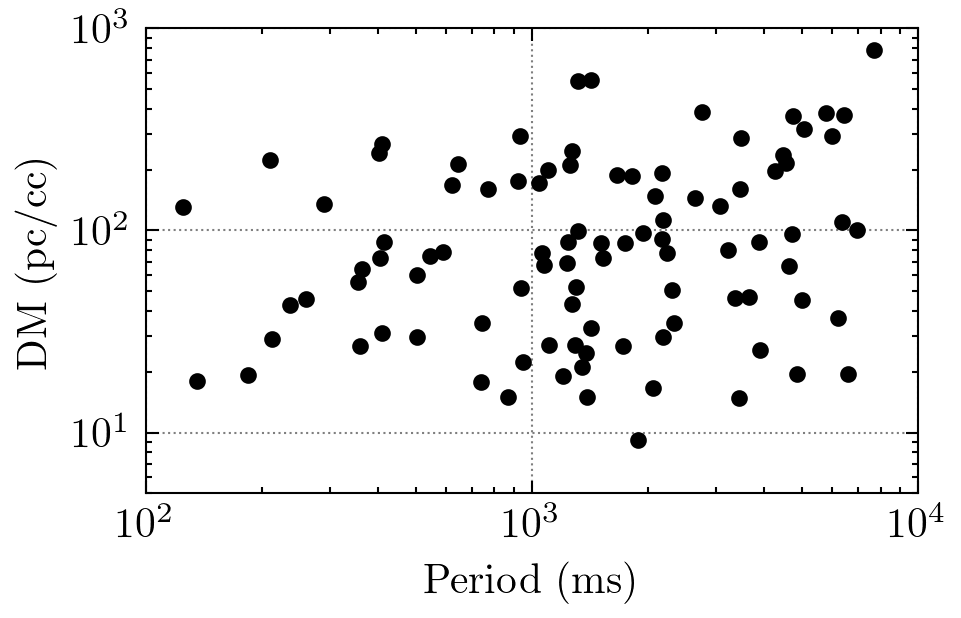}
    \caption{Scatter diagram showing period versus dispersion measure distribution for the sample. Unlike the sample of pulsars, there appears to be no significant observational selection against short period, high DM RRATs.}
    \label{fig:pdm}
\end{figure}

The sample of 335 RRATs currently known, hereafter referred to as the \textsc{RRATalog}\footnote{The \textsc{RRATalog} is freely available and the latest version can be accessed online at \url{https://rratalog.github.io/rratalog}.}, is presented in tabular form in Appendix A. 
In Table A1, for each RRAT we provide the name, spin period ($P$), period derivative ($\dot{P}$), dispersion measure (DM), the sky location in Galactic longitude ($l$) and latitude ($b$), burst rate (${\cal B}$), peak flux density at 1400~MHz ($S_{1400}$), and pulse width measured at 1400~MHz ($W_{1400}$). { For $W_{1400}$, we quote the average of the pulse widths measured at 50\% of the peak intensity.}
Before proceeding to develop models of the RRAT population, we provide various visualisations of this observed sample. Fig.~\ref{fig:sky_distribution} shows the sky distribution of RRATs.  Many sources can be seen along the Galactic plane; this likely reflects that, similar to pulsars, the RRAT number density is higher in the Galactic plane and also that a number of surveys so far have targeted the Galactic plane. We attempt to model these factors in our Monte Carlo simulations detailed below. Fig.~\ref{fig:observed_hist} shows histograms of the observed quantities. With the exception of $P$ and $\dot{P}$ which are discussed in context with the normal pulsars below, we find no statistically significant correlation between any of these four quantities. As an example, Fig.~\ref{fig:pdm} shows the scatter plot between $P$ and DM.

\begin{figure}
    \centering
    \includegraphics[width=\columnwidth]{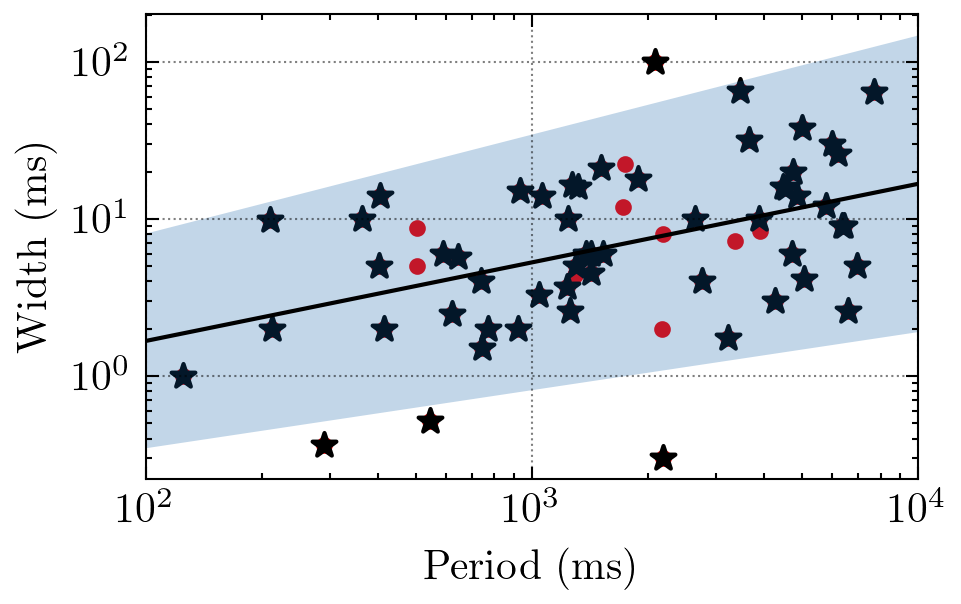}
    \caption{Observed distribution of intrinsic pulse widths and spin periods for the 91 RRATs with measured periods. The black stars correspond to the intrinsic pulse widths derived from observations at 1400~MHz while the red dots correspond to the observations at 350~MHz. The black line shows the fit as described in Eq.~\ref{eq:p_w} with 1$\sigma$ error bars as the blue shaded region.}
    \label{fig:pw}
\end{figure}

As is the case for the pulsar population (see, e.g., LFL06), there exists a correlation between pulse period and pulse width. Since this will be an integral part of our modeling process, we need to derive a period -- pulse width relationship for RRATs. To do this, for each source in the \textsc{RRATalog}, we estimate its intrinsic pulse width
\begin{equation}
    W_\text{int} = \sqrt{W_\text{obs}^2 - t_\text{DM}^2 - t_\text{samp}^2}.
\end{equation}
Here $W_\text{obs}$ is the observed pulse width as tabulated under $W_{1400}$ in Appendix A, $t_\text{DM}$ is the dispersion delay across an individual filterbank channel, and $t_\text{samp}$ is the sampling interval of the survey that found the RRAT. We compute $t_\text{DM}$ using the
published parameters of the appropriate survey and the DM of each RRAT. and the above expression to determine $W_\text{int}$.
The scatter diagram shown in Fig.~\ref{fig:pw} shows the result of this analysis and the mild correlation between
pulse width and period. We model this trend in our simulations below using the simple expression
\begin{equation}
    \label{eq:p_w}
    \log_{10}(W_\text{int}/{\rm ms}) = A \log_{10}(P/{\rm ms}) + B,
\end{equation}
and determine the coefficients $A = 0.49\pm0.13$ and $B = -0.77 \pm 0.42$. This is broadly consistent with the pulse width-period relation found for pulsars \citep{2019MNRAS.485..640J}. Using this relationship, in our simulations we assign each model RRAT an intrinsic pulse width based upon its period.

\begin{figure}
    \centering
    \includegraphics[width=\columnwidth]{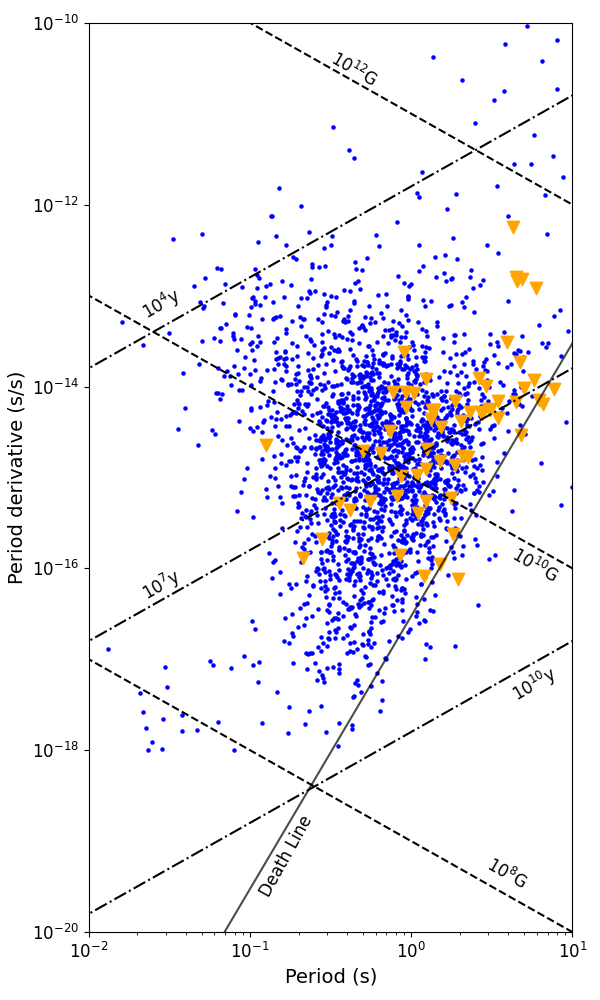}
    \caption{The $P-\dot{{P}}$ diagram showing { canonical} pulsars (dots) and RRATs (triangles). The dashed lines depict  constant magnetic field and the dotted lines show  constant characteristic age. The death line is calculated using Eq.~13 from \citet{1992A&A...254..198B} { and discussed in the text}. }
    \label{fig:ppdot}
\end{figure}

For 105 RRATs, only a few pulses, or sometimes just one, have been detected. In such cases it is not possible to deduce the spin period. Of the 230 other RRATs for which a spin period has been determined, only { 61} RRATs have sufficient detections to enable measurement of both $P$ and $\dot{P}$. These are presented in Table A2 and shown alongside the pulsar population in Fig.~\ref{fig:ppdot}. For such RRATs we present two derived quantities: the surface magnetic field strength ($B_\text{s} = 3.2 \times 10^{19} \sqrt{P \dot{P}}$~G) and the spin-down age ($\tau_c = P/2\dot{P}$). As can be seen, RRATs generally occupy the upper right corner with large periods as compared to the canonical pulsars. The diagram also shows the lines of constant surface magnetic field and spin-down age along with the \citet{1992A&A...254..198B} death line, { the boundary on the $P-\dot{P}$ diagram beyond which radio emission is not predicted to exist, originally dating back to the work of \citet{rs75}. As can be seen, the existence of sources beyond the death line indicates that our understanding of the physical processes involving radio emission is incomplete \citep[for a recent review, see][]{pk22}.}
A two dimensional Kolmogorov--Smirnov (KS) test on the $P$--$\dot{P}$ plane for canonical pulsars and RRATs suggests that their distributions differ with a confidence level of over 99.7\%. The most stark difference seen from a comparison of the observed RRAT and pulsar samples is in their distributions of spin period. For 230 RRATs for which period determinations have been made, the median period is 1.73~s. This is significantly longer than the median period of only 0.66~s for canonical pulsars (which we selected from the current Australia Telescope National Facility (ATNF) catalog as all Galactic pulsars with $\dot{P}>10^{-18}$). Similar conclusions have been drawn from previous studies of the RRAT population, and this period dependence is perhaps partially due to selection effects
\citep{Keane2011,GBNCC2015KA,2022JApA...43...75A}. As we argue later in \S \ref{sec:prrat}, this significant difference
between the two samples shows that RRATs are a better diagnostic of the population of long-period neutron stars than
canonical pulsars alone.

In order to explore whether the RRAT emission properties might depend on spin-down properties, we examined the correlations between these quantities. For the 40 RRATs with measurements of both period derivative and burst rate, we did find a weak positive  correlation between burst rate and age ($r=0.21$), a weak negative correlation between burst rate and inferred surface dipole magnetic field ($r=-0.24$), and no correlation between burst rate and spin-down energy loss rate ($r=0.03$), where $r$ is the Pearson correlation coefficient. This indicates that the detected burst rate is largely determined by other factors aside from intrinsic spin-down properties.

To conclude this overview of the RRATalog, we examined the subset of RRATs lacking timing solutions and compared the distributions of DM, $l$, $b$ and pulse width between the 105 RRATs without periods currently against the 230 RRATs with identified periods. Two-sample KS tests reveal no statistically significant differences in the distributions of DM or pulse width, which suggests that both groups share similar physical characteristics and distances within the Galaxy. While marginal differences were found in the Galactic longitude and latitude  distributions, these are likely attributable to the non-uniform footprints and varying dwell times of the specific sky surveys that discovered them. Consequently, we conclude that the RRATs for which a period has not yet been determined do not represent a distinct class of transients  \citep[such as misidentified extragalactic fast radio bursts;][]{2016MNRAS.459.1360K} but are instead simply the low-repetition tail of the general Galactic RRAT population.

\section{\psrpoppy}
\label{sec:3}
\textsc{PsrPopPy} is a Python-based pulsar population simulation package developed by \citet{Bates2014} to carry out Monte Carlo simulations of the Galactic pulsar population. The package has two modes for generating  populations: a ``snapshot'' mode which generates the present day (static) pulsar population and an ``evolve'' mode which evolves pulsars in time through the Galactic potential and computes their spin-down parameters using a time evolution model. For the snapshot method, the statistical models for the pulsar populations detail the luminosity, spin period and the spatial distributions. For the evolve method, the spin-down model provides an additional period derivative distribution. In this section, we describe an upgrade to the package which we call \psrpoppy, so that it can now perform Monte Carlo simulations of RRATs in the snapshot mode. We defer evolve-mode models of the RRAT population to a subsequent study. { For a recent application of this approach to the pulsar population, see \citet{grp+2024}.} 

For a snapshot model of the RRAT population, we consider distributions of pulse period ($P$), luminosity ($L$) and spatial distribution (radial distance, $R$, and height above the Galactic plane, $z$), from which $N$ RRATs are drawn. Each of these model RRATs is then subject to filters which attempt to mimic the same selection criteria used in the actual pulsar surveys. Next, different pulsar surveys can be applied to this population of $N$ RRATs to simulate the survey yields. The surveys are modelled based on their sky coverage, detection thresholds, telescope gain ($G$), centre frequency, bandwidth ($\Delta f$), frequency and time resolution, number of polarizations recorded ($n_p$), { a} degradation factor { to account for finite-level digitisation} ($\beta$), system temperature $T$ and observation length ($t$). With a given survey, for RRATs inside the sky coverage area, scattering and smearing effects are added to compute the observed pulse width $W_\text{obs}$. For each RRAT with peak flux density $S_\nu$, using a modified version of the pulsar radiometer equation \citep[see, e.g.,][]{1984bens.work..234D}, we compute the signal-to-noise ratio (S/N) at which it would be detected in a periodicity search
\begin{equation}\label{eq:snperiodic}
   \text{S/N}_{\rm periodic} = \frac{S_\nu G \sqrt{n_p t \Delta f }}{\beta T} \sqrt{\frac{P-W_\text{obs}}{W_\text{obs}}}.
\end{equation}
As described below, to follow the detection process to its logical conclusion, we use this ``pulsar survey detection threshold'' to determine whether a source counts as a RRAT in a given survey. { Equation~\ref{eq:snperiodic} represents the best-case scenario for a periodicity search. In practice, long-term variations in the
observing system will make longer-period sources harder to detect in `standard' Fourier-based searches. For this reason,
it is now commonplace to complement periodicity searches with a pipeline using the fast-folding algorithm \citep{1969IEEEP..57..724S} which is less prone
to such variations and coherently sums the pulses over the entire observation \citep[see, e.g.,][]{2020MNRAS.497.4654M}. Improvements in both approaches have been
made in recent years and reprocessing of archival surveys, including those considered here, has led to further pulsar discoveries \citep[see, e.g.,][]{2019MNRAS.483.3673M,2023MNRAS.522.1071S}. Throughout this work, we will assume that the surveys modeled here been sufficiently searched such that Equation~\ref{eq:snperiodic} represents a good
approximation to the S/N achievable in a periodicity search.}

Pulsars are most often  discovered using periodicity searches, though many are also detectable through single-pulse searches. RRATs on the other hand, were discovered  only through single-pulse search techniques and require a different survey threshold model than the one given above. To model their detection, we must account for a RRAT's intrinsic burst rate (${\cal B}$), which is the number of bursts emitted per hour. In our simulations, { as described further in the next section,} $N$ RRATs are drawn with values from $P$, $L$, $R$, $z$ and ${\cal B}$. We compute the
pulse widths using the pulse width--period relationship described in Eq.~\ref{eq:p_w}. To model the scatter in the pulse width--period relation (see Fig.~\ref{fig:pw}), we draw the variables $A$ and $B$ in Eq.~\ref{eq:p_w} from normal distributions using the fitted values and their errors ($A = 0.49\pm0.13$ and $B = -0.77 \pm 0.42$) as the mean and the standard deviations, respectively. As in the earlier study of the pulsar population (LFL06), we do not account for any scintillation of the detected pulses as this is typically not an important factor given the distances to the sources and survey observing frequencies.

We assume, for simplicity, that the RRAT pulse emission is a random process following Poisson statistics. For an observation of length $T_\text{obs}$  and a burst rate $\cal B$, we define $\lambda = \cal B \, T_\text{obs}$ as the expected number of bursts. The probability of detecting $k$ bursts, 
\begin{equation}
    {\cal P} (k) = \frac {\lambda^{k}e^{-\lambda}}{k!}.
\end{equation}
If a RRAT yields zero bursts, it is considered as not detectable by the survey. Otherwise, the amplitude of the busts are drawn from a log-normal distribution with mean value drawn from the $L$ distribution and a standard deviation taken to be $L/\sigma_0$, where $\sigma_0$ is a constant scaling factor. For simplicity, we do not modulate the pulse width in each of the single pulses. For each pulse, following \citet{McLaughlin_Cordes_2003}, we  compute its   signal-to-noise ratio 
\begin{equation}
    \text{S/N}_{\text{single}} = \frac{ S_\text{max} G \sqrt{n_p t \Delta f }}{\beta T}.
\end{equation}
Here $S_\text{max}$ is the peak flux density of the brightest single pulse. If $\text{S/N}_{\text{single}}>\text{S/N}_{\rm periodic}$, { and if $\text{S/N}_{\text{single}}$ is above the threshold $\text{S/N}_{\text{single,min}}$ listed in Table \ref{tab:pulsar_surveys},} the RRAT is considered as detected in the survey and the number of detectable pulses is saved. This number is used to compare the burst rates of observed and modeled detected RRATs for different surveys. The simulation proceeds until the observed number of RRATs are detected in the surveys. 

\begin{table*}
\centering
\caption{Summary of {\it PsrPopPy} model specifications for the pulsar survey parameters considered in this work. From left to right, the surveys considered are: Parkes \citep[PMSURV;][]{pmsurv}, Swinburne intermediate latitude survey \citep[Swin-IL;][]{Edwards}, Swinburne high latitude survey \citep[Swin-HL;][]{Jacoby} and the High Time Resolution Universe mid-latitude survey  \citep[HTRU-mid;][]{HTRU-RRATs, HTRUmid}. We also list specifications for recent, ongoing and future surveys conducted with the Pulsar Arecibo L-Band Feed Array \citep[PALFA;][]{2006ApJ...637..446C}, the Deep Synoptic Array \citep[DSA;][]{2025AAS...24614305W}, the Five Hundred Metre Aperture Spherical Telescope  \citep[FAST;][]{2021RAA....21..107H}, and MeerKAT \citep{2024MNRAS.531.3579T}.}
\begin{tabular}{cccccccccc}
\hline \hline
    & &\multicolumn{8}{c}{Survey} \\
Parameter & Unit & PMSURV & Swin-IL & Swin-HL & HTRU mid & PALFA & DSA & FAST & MeerKAT \\ \hline
{ Degradation factor} & & { 1.25} & { 1.25} & { 1.25} & { 1.07} & { 1.01} & { 1.00} & { 1.00} & { 1.00} \\ 
{ S/N threshold}      & & { 9.0} & { 9.0} & { 9.0} & { 9.0} & { 9.0} & { 9.0} & { 9.0} & { 9.0} \\
Antenna gain & K/Jy & 0.6 & 0.6 & 0.6 & 0.6 & 8.5 & 10 & 16 & 2.8 \\ 
Integration time & s & 2100 & 264 & 264 & 540 & 268 & 900 & 300 & 600 \\ 
Sampling time & $\mu$s & 250 & 125 & 125 & 64 & 64 & 100 & 49 & 74 \\ 
System temperature & K & 25 & 25 & 25 & 23 & 25 & 25 & 25 & 18 \\ 
Centre frequency & MHz & 1374 & 1372 & 1372 & 1352 & 1350 & 1300 & 1250 & 1284 \\ 
Bandwidth & MHz & 288 & 288 & 288 & 340 & 340 & 1300 & 1300 & 776 \\ 
Channel bandwidth & kHz & 3000 & 3000 & 3000 & 390 & 300 & 134 & 244 & 757 \\ 
Beam width & arcmin & 14 & 14 & 14 & 14 & 3.6 & 0.06 & 3 & 1.7 \\ 
Min declination & deg & --90 & --90 & --90 & --90 & 0 & --37 & --14 & --90 \\ 
Max declination & deg & 27 & 27 & 27 & 27 & 38 & 90 & 65 & 40 \\ 
Min Galactic longitude & deg & --100 & --80 & --100 & --120 & 32 & --180 & --180 & --100 \\ 
Max Galactic longitude & deg & 50 & 30 & 50 & 30 & 77 & 180 & 180 & --10 \\ 
Min $|$Galactic latitude$|$ & deg & 0 & 5 & 15 & 0 & 0 & 0 & 0 & 0 \\ 
Max $|$Galactic latitude$|$ & deg & 5 & 15 & 30 & 15 & 5 & 90 & 10 & 5 \\ \hline
\end{tabular}
\label{tab:pulsar_surveys}
\end{table*}

\section{Population Analysis}
\label{sec:4}

We now describe the methods used to generate { a model of} the underlying population of RRATs using \psrpoppy. The central idea is that when such a population  is run through our models of the surveys, the distribution parameters of the detected RRATs match with the distributions of observed RRATs. The model is created using the 55 RRATs detected by four surveys with the Parkes telescope in Australia: the Parkes multibeam pulsar survey \citep{pmsurv}, the high time resolution intermediate survey \citep{HTRU-RRATs, HTRUmid}, and two high latitude surveys \citep{BS10, Jacoby, Edwards}. These surveys have been conducted at L-Band (1.4~GHz) and the parameters used for them in the simulations are summarized in Table \ref{tab:pulsar_surveys}. The primary motivation for selecting this sample is that it represents a well understood observing system { where the results have been thoroughly searched} and does not require assumptions about the spectral index distribution of RRATs. We defer an analysis using a larger sample of RRATs over more surveys for a future paper.

\begin{figure*}
\centering
\includegraphics[width=0.8\textwidth]{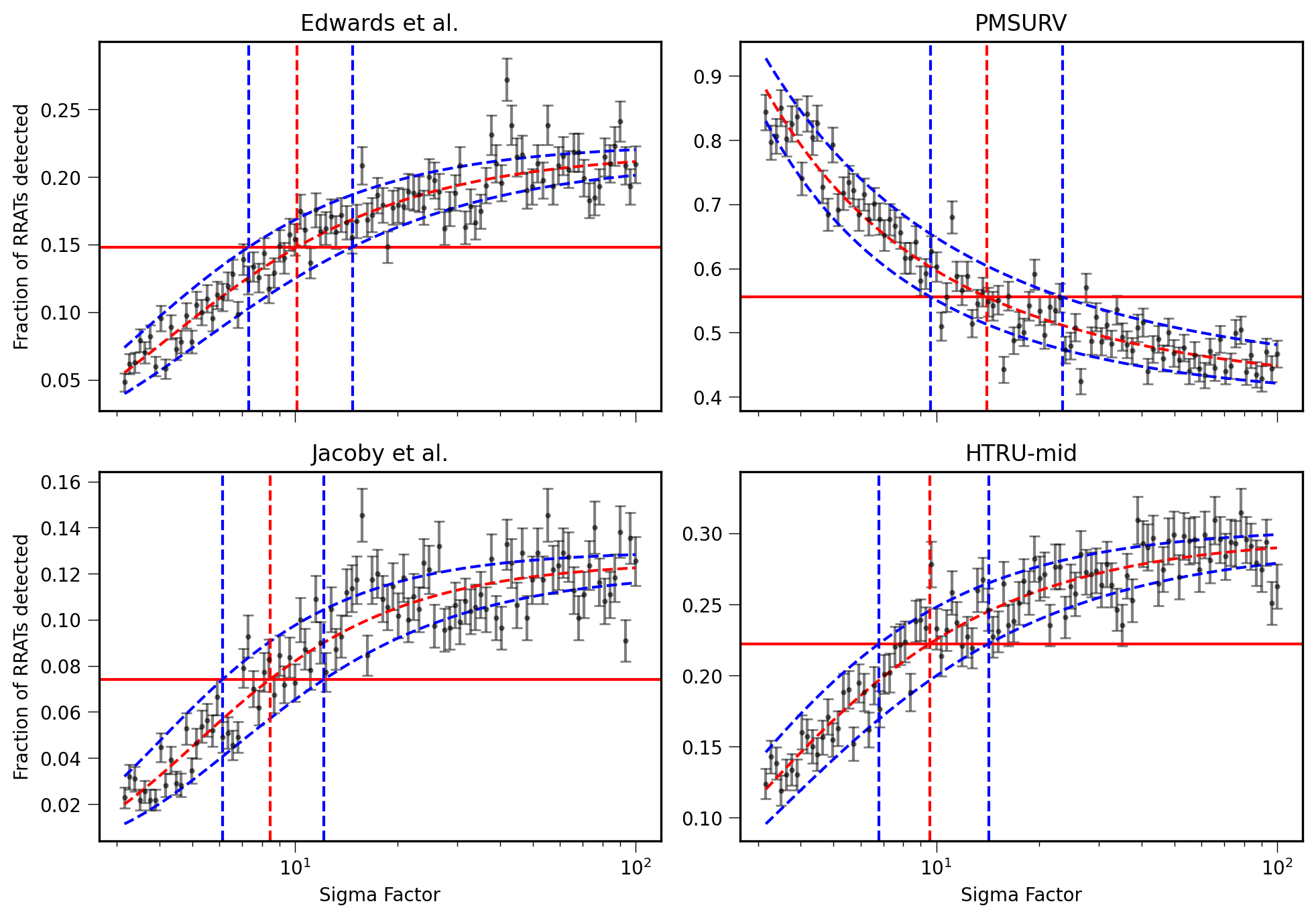}
\caption{The luminosity scaling factor, $\sigma_0$, estimates as determined for each of the four surveys. The error bar on each plot shows the fraction of RRATs detected as a function of the sigma factor. The dashed red curve is the fit to Eq.~\ref{eq:sigma_factor} along with blue dashed curves representing the 68\% confidence intervals. The solid horizontal red line shows the observed fraction of RRATs for the respective surveys. The vertical dashed red line shows estimated $\sigma_0$ value along with blue vertical dashed lines depicting the error region.}
\label{fig:sigma_fix}
\end{figure*}

\begin{table}
	\centering
	\caption{Starting distributions for the underlying population parameters: spin period ($P$), vertical height above the Galactic plane ($z$), Galactocentric radius ($R$), luminosity ($L$) and burst rate (${\cal B}$). As described in the text, the $z$ distribution was ultimately fixed to be an exponential with scale height of 330~pc.}
	\label{tab:bins}
    \begin{tabular}{lcccc}
    \hline
    Parameter   & Unit      & Distribution & Range  & Bins\\
    \hline
    $P$ & s & Uniform & 0.001, 7800 & 8 \\
    $z$ & kpc & Uniform & --1.1, 1.9& 8 \\
    $R$ & kpc & Uniform & 0,12.3 & 8 \\
    $L$ &  mJy kpc$^{2}$& Log--Uniform & 1.5, 4.7 & 8 \\
    ${\cal B}$ & hr$^{-1} $& Log--Uniform & 0.3, 1000&  6\\
    \hline
    \end{tabular}
\end{table}

Following LFL06, we begin with uniformly weighted underlying distributions for the $P$, $z$ and $R$ and log-uniform distributions for $L$ and ${\cal B}$. The parameters and starting values for these distibutions are given in Table~\ref{tab:bins}. We run the simulation until a total of 1,100 RRATs are detected through the surveys mentioned above. This number is 20 times higher than the actual number detected through the surveys in order to minimize statistical fluctuations. The properties of the model-detected population are then compared with the RRATs detected from these surveys by calculating the reduced $\chi^2$ of scaled versions of the distributions in $R$, $L$, $z$ and ${\cal B}$ for the model RRATs when compared to the observed sample.

As detailed in LFL06, { the uniform or log-uniform distributions are simply starting points to sample the input distributions}. { The initial runs of these simulations generally produce a poor match to the
observed sample and result in large $\chi^2$ values. We follow their approach and improve all the distributions by applying correction factors. For each bin, we compute the corresponding correction factor 
\begin{equation}
    C_i = \frac{R_i - M_i}{M_i},
\end{equation}
where $R_i$ and $M_i$ are the number of real and model RRATs observed. These factors are applied to the underlying population to refine the models. For a distribution $X$, the $i$th bin is updated as
\begin{equation}
    X_i^{\text{new}} = X_i + C_i\times X_i.
\end{equation}
Using the updated underlying population, the simulation is repeated until the reduced $\chi^2$  is $\sim$ 1; this typically takes $\sim$15 iterations. We found that, unlike what was seen in LFL06 for the pulsar analysis, our results are relatively insensitive to the scale of the RRAT $z$ distribution and (following LFL06) ended up fixing this distribution to be an exponential with a mean of 330~pc.

\begin{table}
	\centering
	\caption{Fit parameters $a$, $b$ and $c$ defined in Eq.~8 that were used to determine the appropriate luminosity scaling factor, $\sigma_0$ as shown in Fig.~6.}
	\label{tab:fit_table}
    \begin{tabular}{lrrr}
    \hline
    Survey         & \multicolumn{1}{c}{$a$}                & \multicolumn{1}{c}{$b$}                & \multicolumn{1}{c}{$c$}\\
    \hline
    PMSURV         & 0.67 $\pm$ 0.04  & $-1.41$ $\pm$ 0.17 & 0.38 $\pm$ 0.01  \\
    HTRU-Mid       & $-1.29$ $\pm$ 0.13 & 2.36 $\pm$ 0.20  & $-0.52$ $\pm$ 0.01 \\
    Jacoby et al.  & $-3.46$ $\pm$ 0.47 & 2.93 $\pm$ 0.24  & $-0.90$ $\pm$ 0.01 \\
    Edwards et al. & $-2.07$ $\pm$ 0.22 & 2.49 $\pm$ 0.19  & $-0.66$ $\pm$ 0.01 \\
    \hline
    \end{tabular}
\end{table}

To determine the $\sigma_0$ value we carry out the following experiment. We run the above stated algorithm for 100 uniformly log-spaced values of $\sigma_0$ between 3--100. As $\sigma_0$ increases, the width of the log-normal distribution shrinks and starts to look more like a delta function yielding most of the single pulses with approximately same amplitude. We estimate the number of RRATs detected by each survey and compute the fraction of RRATs detected as the number of RRATs detected in the survey divided by 1,100 (the total number of detected RRATs). Fig.~\ref{fig:sigma_fix} shows the fraction of RRATs detected for the four surveys as a function of $\sigma_0$. We then fit a function 
\begin{equation}
    f(\sigma_0) = a \exp{(-b \sigma_0)} + c,
    \label{eq:sigma_factor}
\end{equation}
where the fit parameters for the surveys ($a$, $b$ and $c$) are reported in Table~\ref{tab:fit_table}. The $\sigma_0$ factor is estimated numerically using the Newton-Raphson method to find the the intersection of the fit and the observed fraction (solid red line in Fig.~\ref{fig:sigma_fix}). Fig.~\ref{fig:sigma_avg} shows the $\sigma_0$ from the four surveys. We compute a weighted average of the above and estimate $\sigma_0 = 11 \pm 2$. We adopt a constant $\sigma_0 = 11$ for the simulations.

\begin{figure}
    \centering
    \includegraphics[width=0.9\columnwidth]{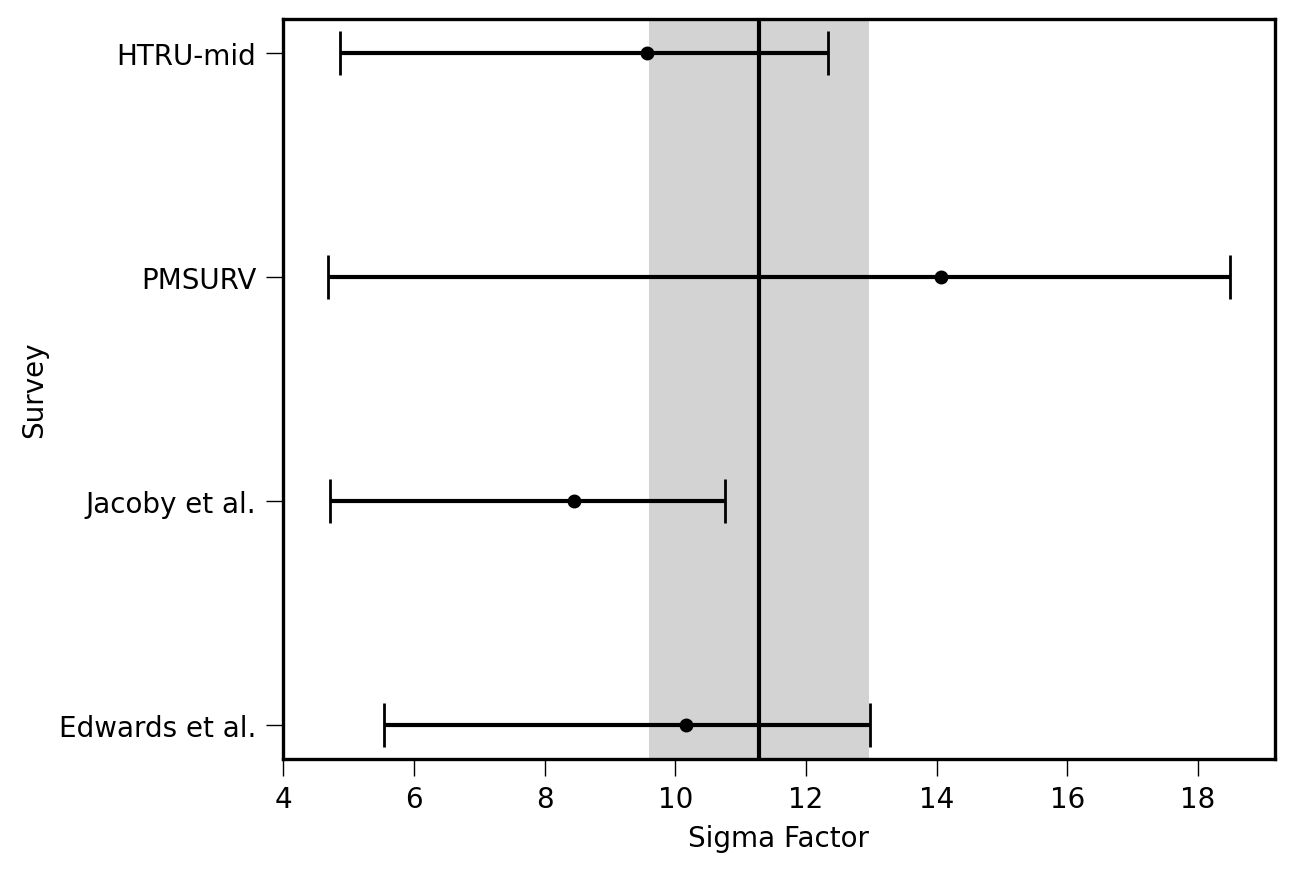}
    \caption{The luminosity scaling factor, $\sigma_0$, estimates from the four surveys. The horizontal error bars show the $\sigma_0$ estimates from the surveys labelled on the y-axis. The black vertical line shows the weighted average along with the error shown in the shaded region.}
    \label{fig:sigma_avg}
\end{figure}

\begin{figure*}
    \centering
    \includegraphics[width=\textwidth]{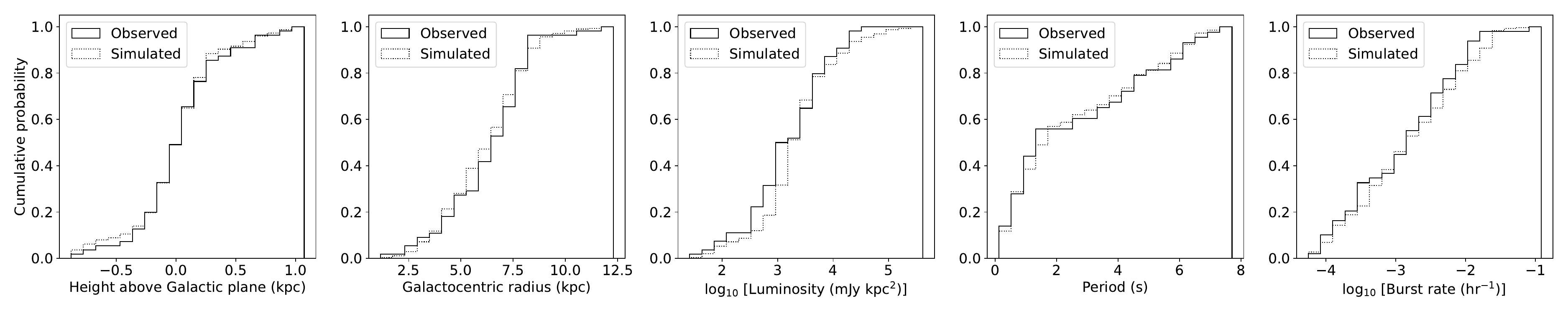}
    \caption{Cumulative density functions for a selection of the observed and derived properties showing the simulated observed RRATs obtained from our optimized population compared to the real observed sample of 55 RRATs used in this study.}
    \label{fig:cdfs}
\end{figure*}

\section{Results}
\label{sec:results}

Using the procedures described above, we obtained an underlying RRAT population that provides an optimal match to the sample of 55 detectable RRATs. Fig.~\ref{fig:cdfs} shows a selection of cumulative density functions from the best-fitting simulated observable population highlighting the excellent agreement between it and the actually observed sample. Our results are best conveyed by fitting smooth functions to the underlying population of RRATs. The various parameters from these distributions are  defined below and their best-fit values and 1$\sigma$ errors are summarized in Table \ref{tab:fit_nos_table}. Fig.~\ref{fig:fitted_hists} shows the distributions along with their best-fitting functional forms.

\begin{table}
	\centering
	\caption{Summary of the best-fitting model parameters to the relevant distributions of our optimized RRAT population model. Mathematical definitions of each parameter are given in \S \ref{sec:results}. Quoted uncertainties represent $1\sigma$ confidence intervals obtained from the covariance matrix of each fit.}
	\label{tab:fit_nos_table}
    \begin{tabular}{lrl}
    \hline
    Parameter       & Value     & Unit\\
    \hline
    $\rho_0$   & 80 $\pm$ 10      & kpc$^{-2}$\\
    $A$   & 2.6 $\pm$ 0.6       & \\
    $B$   & 5.0 $\pm$ 1.0       & \\
    $\alpha$   & $-1.34 \pm 0.05$    & \\
    $C$   & 7.0 $\pm$ 0.2       & \\
    $D$   & 1000 $\pm$ 400     & \\
    $E$   & 6 $\pm$ 3           & s\\
    $F$   & 7000 $\pm$ 500      & \\
    $G$   & 0.44 $\pm$ 0.10       & s\\
    $H$   & 0.7 $\pm$ 0.1       & s\\
    $I$   & $0.014 \pm 0.002$& \\
    ${\mathcal B}^*$   & $0.40 \pm 0.03$ & min$^{-1}$\\
    $J$   & $-1.6 \pm 0.1$& \\
    \hline
    \end{tabular}
\end{table}

In Fig.~\ref{fig:fitted_hists}a, following LFL06, we compute the radial surface density of RRATs, $\rho$, in each bin of Galactocentric radius, $R$, and fit this to a gamma distribution where
\begin{equation}\label{equ:rho}
    \rho(R) = \rho_0 \bigg[ \frac{R}{R_\odot} \bigg]^A \exp{\bigg[ -B \left(\frac{R - R_\odot}{R_\odot}\right) \bigg] }.
\end{equation}
Here $R_\odot$ is the distance of the Sun from the Galactic centre and is taken to be 8.5~kpc and $\rho_0$, $A$ and $B$ are free parameters, where $\rho_0$ represents the local surface density of RRATs (i.e., at $R=R_\odot$). 
To compute the population size, $N$, we integrate Eqn.~\ref{equ:rho} over all values of $R$ with cylindrical symmetry
in the azimuthal angle $\phi$ to find
\begin{equation}
N = \int_{0}^{2\pi} {\rm d}\phi \int_{0}^{\infty} \rho(R) R \, {\rm d}R  = 2\pi \rho_0 R_\odot^2 \frac{e^B \Gamma(A+2)}{B^{A+2}},
\end{equation}
where, as usual, $\Gamma(n) = \int_0^\infty x^{n-1}e^{-x}{\rm d}x$. Taking into account the uncertainties in  $A$ and $B$, the
result is $N = 44000 \pm 8000$. We note that our analysis is insensitive to RRATs with luminosities below the faintest
detected in our sample, $L_{\rm min}\simeq 10$~mJy~kpc$^2$. This result therefore corresponds to the population of potentially observable RRATs (i.e.,~those sources beaming towards Earth) with luminosities above $L_{\rm min}$. We discuss the implications
of this result in the context of the pulsar population as a whole in Section \ref{sec:6}.

For the luminosity function, for luminosities  $\gtrsim$~30~mJy~kpc$^2$, Fig.~\ref{fig:fitted_hists}b is
well described by a power law in which 
\begin{equation}
    \log N = \alpha \log L + C,
\end{equation}
where $\alpha=-1.34$ is the slope of the differential distribution and $C$ is a normalizing constant. To integrate this
function over different luminosity ranges, we note that
    ${\rm d}N = (10^C/\ln 10) L^{\alpha-1} {\rm d}L$. For $\alpha<0$, which is the case here, this function integrates to give
\begin{equation}\label{eq:nlmin}
    N(L>L_{\rm min}) = -\frac{10^C L_{\rm min}^\alpha}{\alpha \ln 10} \simeq 34000 \left(\frac{L_{\rm min}}{{\rm 30~mJy~kpc}^2}\right)^{-1.34},
\end{equation}
which is valid for $L_{\rm min}\gtrsim$~30~mJy~kpc$^2$.
We discuss the form of the RRAT luminosity function further in Section \ref{sec:6}.

For the period distribution shown in Fig.~\ref{fig:fitted_hists}c, we found that a satisfactory fit was obtained using a sum of exponential and Gaussian functions so that
\begin{equation}\label{eq:pdist}
    N(P) = D \exp{\bigg[ -\frac{P}{E}\bigg]} + F \exp{ \bigg[-\frac{1}{2} \bigg( \frac{P - G}{H} \bigg)^2 \bigg] },
\end{equation}
where the fitted parameters are $D$, $E$, $F$, $G$ and $H$, respectively.
Finally, as shown in Fig.~\ref{fig:fitted_hists}d, the underlying RRAT burst rate distribution for burst
rates ${\mathcal B} \gtrsim 6 \times 10^{-5}$~hr$^{-1}$ is well described as a power law with an exponential cut-off. Following \citet{1976ApJ...203..297S}, this is commonly referred to in astronomy as the Schechter function. We use the function in this context to characterize the burst rate distribution
\begin{equation}
    N({\cal B}) = I \bigg[ \frac{\cal B}{\cal B^*} \bigg]^J \exp \bigg[ - \frac{\cal B}{\cal B^*}\bigg],
\end{equation}
where $I$ is a scaling factor, $\cal B^*$ is the characteristic burst rate and $J$ is the power law index.

\begin{figure*}
    \centering
    \includegraphics[width=\textwidth]{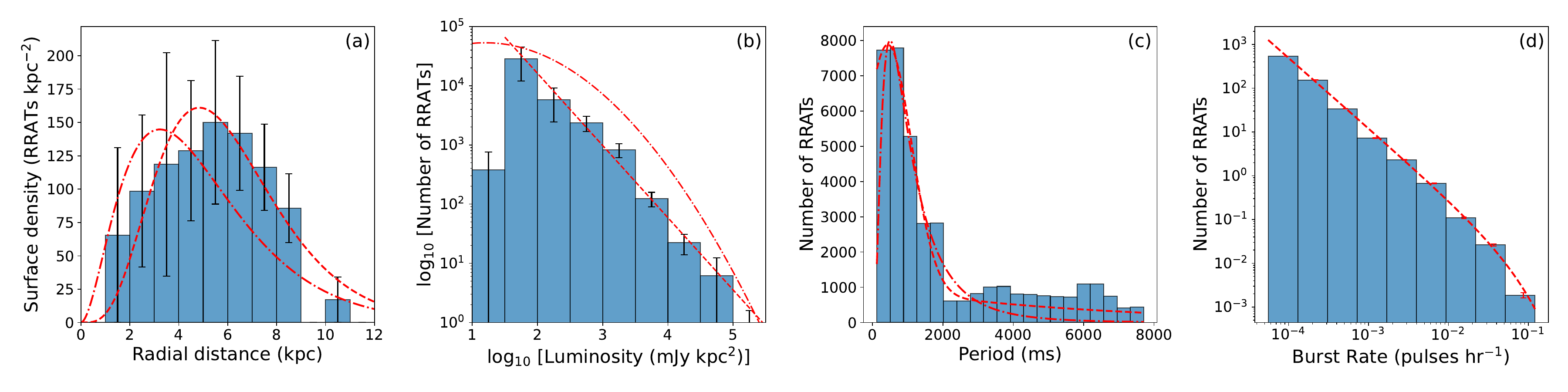}
    \caption{Model parameter distributions and best-fitting functions (dashed lines; see \S5 for details) for the underlying distribution of: surface density as a function of Galactocentric radius (a), luminosity (b), period (c) and burst rate (d). The error bars shown are based on the statistics of the observed sample (i.e., fractional errors of $1/\sqrt{N}$ derived from the appropriate bin in the observed sample). The dash-dotted line shown for the surface density and period distributions (panels a and c) represent the functional form for the canonical pulsar population found by LFL06. As discussed in the text, the dash-dotted line in panel b is a version of the log-normal luminosity function found by FK06 scaled to account for the difference between peak and mean luminosity.}
    \label{fig:fitted_hists}
\end{figure*}
\label{sec:5}

\section{Discussion}
\label{sec:6}

We now compare our results to those found for the pulsar population by LFL06 and earlier
results for RRATs found by \citet{McLaughlin2006}, \citet{Keane2011} and \citet{2008MNRAS.391.2009K}. We also
confront our results with current and future RRAT surveys.

\subsection{The Galactocentric radial distribution of RRATs}

Both the pulsar and RRAT distributions can be approximated by the { functional} form of the radial
density profile given in Eqn.~\ref{equ:rho}. 
As shown in Fig.~\ref{fig:fitted_hists}a, where we compare the form of Eqn.~\ref{equ:rho} found by LFL06 (dashed-dotted line) with our fit to the underlying RRAT population (dashed line), the shapes of the two distributions are
qualitatively similar. The fact that the RRAT distribution prefers higher $R$ values than the pulsar
distribution found by LFL06 most likely reflects the difficulties in finding RRATs in the inner Galaxy by the
surveys considered here. 
For pulsars, LFL06 showed that the form of the radial
density of the underlying population is strongly dependent on the distribution of free electrons used in the simulations. { Throughout this work, we adopted
the electron density model of \citet{ymw16}. LFL06 used the \cite{ne2001} model. Some of the differences shown here could reflect the underlying form 
of those two distributions.}
In particular, different radial
density profiles in the inner Galaxy could be obtained by choosing different electron density distributions in the simulations. For our case, where the sample size of RRATs is much smaller than 
the currently known pulsars, our modeling of the radial density can only probe $R>3$~kpc. { Future studies using the recently developed NE2025 electron density model \citep{2026arXiv260211838O} are now strongly encouraged.}
We discuss the prospects
for current and future RRAT surveys to sample the inner Galaxy later in \S \ref{sec:other}. The simplest conclusion to draw from our results with the earlier
findings of LFL06 is that pulsars and RRATs have a common radial distribution function. The relative sizes of the two populations requires 
a consideration of their luminosity functions which we discuss next.

\subsection{The RRAT luminosity function}\label{sec:sizelf}

As with any astronomical population, quoting the number of sources is heavily dependent on the choice of minimum luminosity, $L_\text{min}$. For RRATs, lower values of $L_\text{min}$ lead to simulation of a large number of low-luminosity sources which are seldom detected in our model surveys. As can be seen in Fig.~\ref{fig:fitted_hists}b, the RRAT luminosity function follows a power law
with slope $\alpha \simeq -1.3$ for luminosities above $\log_{10} L_{\rm min}=1.5$, i.e. $L_\text{min} \simeq  30$~mJy~kpc$^2$.
Using Eq.~12 with the best-fit parameters for $\alpha$ and $C$ given in Table 3, we find $N = 34000 \pm 1600$~RRATs with $L>L_\text{min}=30$~mJy~kpc$^2$ beaming towards us (hereafter referred to as ``potentially observable'' RRATs).

In the RRAT discovery paper, \citet{McLaughlin2006} conducted an analysis in which they infer a population of 16,000 RRATs above 25~mJy~kpc$^2$, somewhat smaller than our determination. The main difference between these two results is our development of a more detailed treatment of the burst rate of RRATs than was available in the earlier study. As can be seen in Fig.~8, the distribution of burst rates required to build a self-consistent population model spans three orders of magnitude. In addition, as discussed further below, we also find evidence for a much steeper RRAT luminosity function,
$\alpha=-1.34$, compared to $\alpha=-1$ assumed by \citet{McLaughlin2006}. In summary, while our results imply a larger population than previously thought, we believe that they more accurately account for the observational progress that has been made since 2006.

To compare our result with canonical pulsars, we need to scale RRAT luminosities to account for the difference between peak intensity values measured for RRATs versus period-averaged ones for canonical pulsars. \citet{2006ApJ...643..332F}, hereafter FK06, estimated
that 120,000 potentially observable pulsars exist in the Galaxy with a log-normal distribution of luminosities with a
mean in log $L$ of --1.1 and a standard deviation of 0.9.
For a top-hat pulse, the mean flux density is simply the peak flux density multiplied by the pulse duty cycle. For the sample of 51 RRATs for
which both periods and pulse widths are available in the RRATalog, we find the median duty cycle to be 0.5\%. Adopting this value, { a peak luminosity of 30~mJy~kpc$^2$ corresponds to a mean luminosity of 0.15~mJy~kpc$^2$. Our results therefore} imply  $34000 \pm 1600$ RRATs with {\it mean luminosities} $\gtrsim 0.15$~mJy~kpc$^2$. To this luminosity limit, the RRAT population is about three-quarters the size of the canonical radio pulsar population. This result can be seen  in Fig.~\ref{fig:fitted_hists}b where we contrast our RRAT luminosity function with a scaled version of the 
FK06 luminosity function which predicts about 45,000 canonical pulsars above the same luminosity limit.

While the exact form of the RRAT luminosity function below $L_\text{min}$ is not well constrained by our analysis, it is clear from Fig.~\ref{fig:fitted_hists}b that
there is a turnover in the function at around 25~mJy~kpc$^2$. We verified this by { boosting the contribution of the lowest luminosity bin} in our simulations and it was found to significantly over predict the detection of RRATs with low flux densities compared to the observed sample. Further analyses of more sensitive RRAT surveys should be 
able to yield better constraints below the current limit. We note for now that, given the likely turnover in the RRAT
luminosity function below $L=30$~mJy~kpc$^2$, the total population of potentially observable RRATs may not be more
than a factor of two higher than the number we have determined for luminosities greater than $30$~mJy~kpc$^2$. Based on this line of reasoning, we conservatively predict that the maximum number of potentially observable RRATs in the Galaxy may not exceed 70,000. This is significantly smaller than the total number of 120,000 potentially observable canonical pulsars inferred by FK06 in their analysis. We revisit these numbers later after correcting the RRAT population for period-dependent beaming effects in the next section.

\subsection{The RRAT period distribution}\label{sec:prrat}

Carrying out a Kolmogorov-Smirnov test on the period distribution found in Fig.~\ref{fig:fitted_hists}c and the sample
of RRATs detected by the Parkes surveys considered in this analysis, we find no evidence that the distributions are different. This finding
is reasonable, given that RRATs are detected initially through their single pulses, their detection thresholds are based on 
radiometer noise considerations over the pulse widths and should not discriminate between different periods. When we look at
the larger sample shown in Fig.~\ref{fig:observed_hist}, however, we note that the observed RRATs have a significantly enhanced
tail. The distribution in Fig.~\ref{fig:observed_hist} is well fitted by a log-normal distribution with a median of 1890~ms 
and a mode of 920~ms. As mentioned earlier, this distribution is significantly different from that seen for the canonical pulsar population. This difference, which is particularly stark when $P>1$~s, is highlighted by the cumulative distribution of periods for RRATs alongside the rest of the pulsars in the ATNF catalogue in Fig.~\ref{fig:Pcdf}. We  interpret the long tail in the RRAT period distribution as being 
a better reflection of the underlying period distribution of rotation-powered neutron stars in general. The dearth of canonical pulsars with longer periods compared to RRATs  likely
reflects the difficulties in detecting them as periodic sources in that region due to the presence of low-frequency noise in 
the data acquisition systems \citep[see, e.g.,][]{2015ApJ...812...81L,2016MNRAS.459.1104L}.

\begin{figure}
    \centering
    \includegraphics[width=0.48\textwidth]{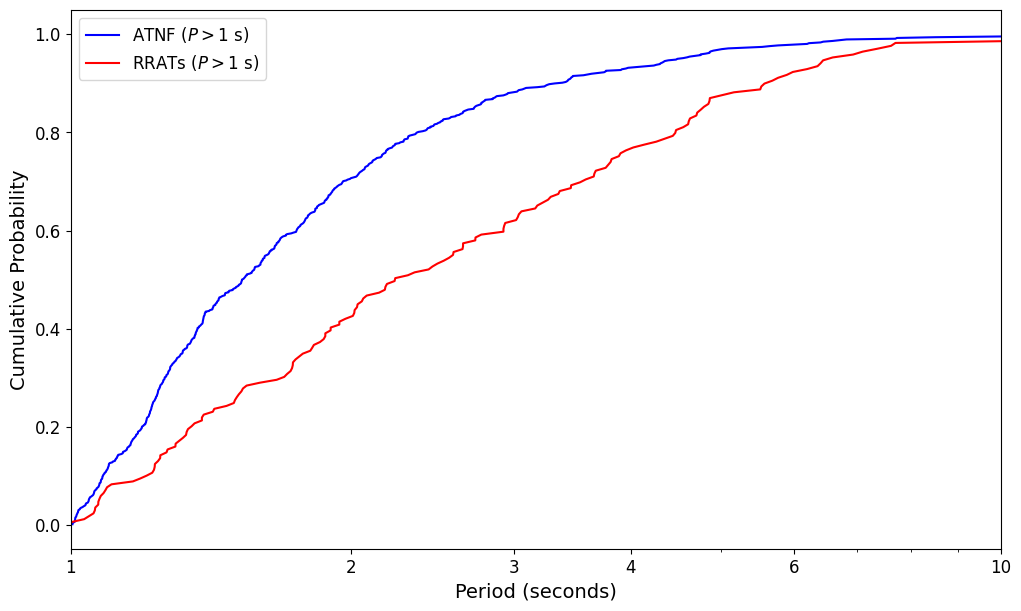}
    \caption{Cumulative distribution of Galactic pulsars with $P>1$~s and $\dot{P}>10^{-18}$ taken from the ATNF catalogue compared to the RRATs. There is a significantly higher fraction of high $P$ objects in the RRAT sample.}
    \label{fig:Pcdf}
\end{figure}

The presence of long-period sources in the potentially observable RRAT population, as revealed by our analysis, coupled
with current insights into the beaming fraction of pulsars, highlights an important issue in neutron star demography. Since it is generally accepted that longer period pulsars have smaller beaming fractions \citep[see, e.g.,][]{1998MNRAS.298..625T}, a natural consequence of the long-tail in our model RRAT period distribution is that the true neutron star period distribution has a larger number of sources at long periods than one might initially expect. In addition to recent discoveries of radio pulsars with periods as long as 76~s \citep{2022NatAs...6..828C}, imaging surveys currently being carried out to detect long-period radio transients \citep{2025arXiv251110785M} have excellent prospects at better constraining the true distribution of 
spin periods of slowly rotating neutron stars. Further progress using the RRAT population in future should be possible using 
time-dependent models to account for the physics of neutron star spindown as opposed to the simpler snapshot modeling presented in this work.

\subsection{The Galactic  population and birth rate of RRATs}

Taking our best-fitting RRAT period distribution from our study based on the Parkes sample and applying the \citet{1998MNRAS.298..625T} beaming model, we find the mean beaming correction over the RRAT period distribution to be around 6.1. This boosts the potentially observable population with $L>30$~mJy~kpc$^2$ discussed in \S \ref{sec:sizelf} to lead us to estimate a total population of $(2.1 \pm 0.1) \times 10^5$ RRATs with peak luminosities $L>30$~mJy~kpc$^2$ in the Milky Way. As discussed in  \S~\ref{sec:sizelf}, this peak luminosity is equivalent to an equivalent canonical pulsar mean luminosity of 0.1~mJy~kpc$^2$. A direct estimate of the birth rate required to sustain this population is challenging. Unlike the canonical pulsar population, where period derivatives are available for the vast majority of sources, the number of RRAT timing solutions is still relatively small and a ``pulsar current analysis'' \citep{1981MNRAS.194..137P,1981JApA....2..315V} as applied to the pulsar population by LFL06 is currently not feasible for RRATs. 

We estimate the RRAT birth rate using the currently available RRAT characteristic ages in Table \ref{tab:RRATspindown} which have a mean 
value of 7~Myr. Assuming a radio lifetime of  twice this value, we infer a rough birth rate of $2\times10^5/(14~{\rm Myr})\simeq1.5$~RRATs per century with peak luminosities $L>30$~mJy~kpc$^2$. Following \citet{2008MNRAS.391.2009K}, a more robust estimate of the birth rate is to assume that RRATs and pulsars have similar
active lifetimes and, therefore, the relative numbers of RRATs and pulsars correspond to their relative birth rates.
As discussed above, based on our observations of a turnover in the RRAT luminosity function, it is unlikely that the total RRAT population exceeds $4\times 10^5$. When compared to the total number of radio-loud pulsars
in the Galaxy estimated by FK06 of around a million, our results suggest that the birth rate of
RRATs are less than half of the total pulsar birth rate. Taking the pulsar birth rate of $2.8 \pm 0.1$ pulsars per century determined by FK06, we estimate the total RRAT birth rate to be $\lesssim$1.4~RRATs per century. 
In summary, while { the uncertainties involved in the above estimates are significant,} our analysis suggests that RRATs do not appear to be the dominant source of radio-loud neutron stars in the Galaxy. As discussed
by \citet{2008MNRAS.391.2009K}, there is significant tension between the birth rate of Galactic
neutron stars and core collapse supernovae. { However, this tension appears mostly due to uncertainties in the birth rates of X-ray dim isolated neutron stars and magnetars \citep[see Fig.~2 of][]{2008MNRAS.391.2009K} rather than RRATs.}

\subsection{Other surveys}\label{sec:other}

The analysis presented in this paper is based on RRAT samples detected by the Parkes surveys. Since these were conducted, several surveys with significantly higher sensitivity have been published. In this section, we confront our population model with these results. 

First, we applied our model to the Arecibo L-band Feed Array
(PALFA) survey. As detailed in Table 1, PALFA surveyed the northern Galactic plane with much greater sensitivity at than the Parkes surveys \citep{2006ApJ...637..446C}. PALFA has been highly successful in discovering intermittent sources, with 20 RRATs attributed to the survey\footnote{For a list of PALFA discoveries, see \url{https://palfa.nanograv.org}.} Using the parameters given in Table 1, our model predicts 40 RRATs from PALFA. This suggests that the current sample of known PALFA RRATs may be significantly incomplete, likely due to the challenges of interference excision \citep{2015ApJ...812...81L} and the limited number of confirmation observations for the faintest candidates now that PALFA is no longer in operation \citep{2022ApJ...924..135P}.

A significant fraction of the currently observed sample has been found by the FAST Galactic Plane Snapshot Survey \citep[FAST-GPPS;][]{2021RAA....21..107H} which discovered 107 RRATs \citep{2025RAA....25a4001H} and redetected 48 previously known sources \citep{2023RAA....23j4001Z}. Using the system parameters given by \citet{2021RAA....21..107H}, and an approximation of the survey area described in Fig.~1 of 
\citep{2025RAA....25a4001H}, we predict a total of $\sim 100$ sources in
this survey. The discrepancy between our model and the FAST-GPPS yield—where the model underpredicts the observed sample—likely reflects the sensitivity of the GPPS to a burgeoning population of low-luminosity sources that are poorly constrained by the less sensitive Parkes surveys, as discussed in \S\ref{sec:sizelf}.

We also carried out similar modeling for the MeerKAT  Galactic Plane Survey \citep[MMGPS-L;][]{2023MNRAS.524.1291P}. By adopting the survey parameters and accounting for the coherent beam sensitivity profiles discussed in the correspondence with the MeerKAT team, we applied our population model to the MMGPS-L survey area. We predict that this survey should detect approximately 40 RRATs. This figure represents a prediction of our best-fit model to be tested by future work, as the final tally of new and redetected sources from the survey is still being finalized. Our predicted yield is consistent with early returns from the survey, demonstrating our model's utility in providing benchmarks for upcoming transient searches.

The next decade will see a transformative increase in the RRAT census with the advent of the Deep Synoptic Array \citep[DSA;][]{2021AAS...23731604R}. While the DSA will conduct a continuous commensal transient search, its dedicated pulsar survey is particularly well-suited for RRAT discovery. Occupying approximately 65\% of the total observing time over five years, this survey will cover the entire sky visible to the array (declinations $\delta > -37^\circ$) 16 times with 15-minute integrations. Operating in the 0.7--2~GHz band, using the parameters summarized in Table 1, we predict that the DSA pulsar search will detect approximately 3,600 RRATs. This order-of-magnitude increase in the known population will provide an unprecedented sample to probe the luminosity function turnover identified in Section~\ref{sec:sizelf}. { The main reasons for such a large haul are the high antenna gain of the DSA coupled with the strategy of multiple passes of the sky. While we have not run detailed models of future surveys with the Square Kilometer Array (SKA), the proposed designs for SKA1-Mid summarized by \citet{2025OJAp....854256K} suggest antenna gains of order 5~K/Jy at 1~GHz will be achievable. While not as sensitive as the DSA will be in the northern sky, it is clear that initial SKA surveys also have excellent prospects for detecting significant numbers of RRATs in the southern hemisphere and probing their Galactic distribution.}

\section{Conclusions}
\label{sec:7}

In this paper, we have presented an updated catalogue of 335 RRATs and utilized a modified version of the population synthesis package, \psrpoppy, to model their Galactic distribution. Our results provide a robust match to the observed sample from four Parkes surveys and allow for a comprehensive estimation of the underlying RRAT population. Our key findings are summarized below.
\begin{itemize}
\item The radial density profiles for RRATs appear to be similar to those found for canonical pulsars.
\item We argue that the period distribution of RRATs in the range $P>1$~s may be a better proxy for the underlying neutron star period distribution than canonical pulsars.
\item We estimate that there are $34000 \pm 1600$ RRATs beaming towards Earth with peak luminosities above 30 mJy kpc$^2$.
\item The RRAT luminosity function follows a power law with a steep slope of $\alpha \approx -1.34$, but shows a significant turnover at lower luminosities.
\item In the luminosity range where both populations are well-sampled, the potentially observable population of RRATs appears to be about three-quarters that of the
population of potentially observable canonical pulsars.
\item After correcting for beaming effects, we find a total Galactic RRAT population of  $\le { 400,000}$, which corresponds to a birth rate of $\le 1.4$ RRATs per century.
\end{itemize}

The simulation results indicate that RRATs are a dominant component of the Galactic neutron star population, significantly outnumbering canonical pulsars at high peak luminosities. Our analysis of the underlying RRAT period distribution, which is skewed toward longer periods compared to those of canonical pulsars, suggests that RRATs are a more evolved population. While the $P-\dot{P}$ distribution further implies that many RRATs are older and approaching the pulsar death line, the current sample with measured period derivatives remains small, and selection effects strongly favor the detection of longer-period sources. The identified turnover in the luminosity function at $L \approx 30 \text{ mJy kpc}^2$ suggests that while RRATs are numerous, the total number of potentially observable sources is likely below 70,000. These findings suggest that the RRAT state may be a common, long-lived phase for aging neutron stars as their steady emission fades or becomes extremely intermittent.

Future surveys with high-sensitivity instruments like FAST and MeerKAT will be crucial in sampling the inner Galaxy and refining the Galactocentric radial distribution model, which is currently limited by the sensitivity of existing surveys for $R < 3 \text{ kpc}$. Such observations will determine if the common radial distribution between pulsars and RRATs holds true across the entire Galaxy. Future work will involve the implementation of ``evolve-mode'' simulations in \psrpoppy\,to study the temporal evolution and spin-down parameters of the RRAT population. Continued timing observations are essential to increase the sample of RRATs with measured period derivatives, which will provide deeper insights into their evolutionary relationship with the broader neutron star population and the findings from upcoming facilities.
\section*{Acknowledgements}

We thank James Turner and Ben Stappers for useful discussions { and the referee for a number of insightful comments on an earlier version of this manuscript. Both E.F.L. and M.A.M. are supported by the award award AST-2009425 from the National Science Foundation (NSF). M.A.M. is also
supported by NSF Physics Frontiers Center award PHYS2020265. D.R.L. and M.A.M. acknowledge support from the Eberly family during the final stages of the preparation of the manuscript.}

\section*{Data Availability}

The up-to-date catalogue of 335 RRATs used in this study, the \textsc{RRATalog}, is available online at \url{https://github.com/rratalog/rratalog}. The software used for the population synthesis modeling, \psrpoppy, is an open-source package available on GitHub. Any other data products or simulation results from the Monte Carlo analysis are available from the authors upon reasonable request.


\begin{thebibliography}{}
\makeatletter
\relax
\def\mn@urlcharsother{\let\do\@makeother \do\$\do\&\do\#\do\^\do\_\do\%\do\~}
\def\mn@doi{\begingroup\mn@urlcharsother \@ifnextchar [ {\mn@doi@} {\mn@doi@[]}}
\def\mn@doi@[#1]#2{\def\@tempa{#1}\ifx\@tempa\@empty \href {http://dx.doi.org/#2} {doi:#2}\else \href {http://dx.doi.org/#2} {#1}\fi \endgroup}
\def\mn@eprint#1#2{\mn@eprint@#1:#2::\@nil}
\def\mn@eprint@arXiv#1{\href {http://arxiv.org/abs/#1} {{\tt arXiv:#1}}}
\def\mn@eprint@dblp#1{\href {http://dblp.uni-trier.de/rec/bibtex/#1.xml} {dblp:#1}}
\def\mn@eprint@#1:#2:#3:#4\@nil{\def\@tempa {#1}\def\@tempb {#2}\def\@tempc {#3}\ifx \@tempc \@empty \let \@tempc \@tempb \let \@tempb \@tempa \fi \ifx \@tempb \@empty \def\@tempb {arXiv}\fi \@ifundefined {mn@eprint@\@tempb}{\@tempb:\@tempc}{\expandafter \expandafter \csname mn@eprint@\@tempb\endcsname \expandafter{\@tempc}}}

\bibitem[\protect\citeauthoryear{{Abhishek}, {Malusare}, {Tanushree}, {Hegde}  \& {Konar}}{{Abhishek} et~al.}{2022}]{2022JApA...43...75A}
{Abhishek} {Malusare} N.,  {Tanushree} N.,  {Hegde} G.,   {Konar} S.,  2022, \mn@doi [Journal of Astrophysics and Astronomy] {10.1007/s12036-022-09862-3}, \href {https://ui.adsabs.harvard.edu/abs/2022JApA...43...75A} {43, 75}

\bibitem[\protect\citeauthoryear{Bates, Lorimer, Rane  \& Swiggum}{Bates et~al.}{2014}]{Bates2014}
Bates S.~D.,  Lorimer D.~R.,  Rane A.,   Swiggum J.,  2014, \mn@doi [MNRAS] {10.1093/mnras/stu157}, 439, 2893

\bibitem[\protect\citeauthoryear{{Bezuidenhout} et~al.,}{{Bezuidenhout} et~al.}{2022}]{2022MNRAS.512.1483B}
{Bezuidenhout} M.~C.,  et~al., 2022, \mn@doi [\mnras] {10.1093/mnras/stac579}, \href {https://ui.adsabs.harvard.edu/abs/2022MNRAS.512.1483B} {512, 1483}

\bibitem[\protect\citeauthoryear{{Bhattacharya}, {Wijers}, {Hartman}  \& {Verbunt}}{{Bhattacharya} et~al.}{1992}]{1992A&A...254..198B}
{Bhattacharya} D.,  {Wijers} R. A.~M.~J.,  {Hartman} J.~W.,   {Verbunt} F.,  1992, \aap, \href {https://ui.adsabs.harvard.edu/abs/1992A&A...254..198B} {254, 198}

\bibitem[\protect\citeauthoryear{{Boyles} et~al.,}{{Boyles} et~al.}{2013}]{2013ApJ...763...80B}
{Boyles} J.,  et~al., 2013, \mn@doi [\apj] {10.1088/0004-637X/763/2/80}, \href {https://ui.adsabs.harvard.edu/abs/2013ApJ...763...80B} {763, 80}

\bibitem[\protect\citeauthoryear{{Burke-Spolaor}}{{Burke-Spolaor}}{2013}]{BS13}
{Burke-Spolaor} S.,  2013, in {van Leeuwen} J.,  ed.,  IAU Symposium Vol. 291, Neutron Stars and Pulsars: Challenges and Opportunities after 80 years. pp 95--100 (\mn@eprint {arXiv} {1212.1716}), \mn@doi{10.1017/S1743921312023277}

\bibitem[\protect\citeauthoryear{{Burke-Spolaor} \& {Bailes}}{{Burke-Spolaor} \& {Bailes}}{2010}]{BS10}
{Burke-Spolaor} S.,  {Bailes} M.,  2010, \mn@doi [\mnras] {10.1111/j.1365-2966.2009.15965.x}, \href {http://adsabs.harvard.edu/abs/2010MNRAS.402..855B} {402, 855}

\bibitem[\protect\citeauthoryear{{Burke-Spolaor} et~al.,}{{Burke-Spolaor} et~al.}{2011}]{HTRU-RRATs}
{Burke-Spolaor} S.,  et~al., 2011, \mn@doi [\mnras] {10.1111/j.1365-2966.2011.18521.x}, \href {http://adsabs.harvard.edu/abs/2011MNRAS.416.2465B} {416, 2465}

\bibitem[\protect\citeauthoryear{{Caleb} et~al.,}{{Caleb} et~al.}{2022}]{2022NatAs...6..828C}
{Caleb} M.,  et~al., 2022, \mn@doi [Nature Astronomy] {10.1038/s41550-022-01688-x}, \href {https://ui.adsabs.harvard.edu/abs/2022NatAs...6..828C} {6, 828}

\bibitem[\protect\citeauthoryear{{Chen} et~al.,}{{Chen} et~al.}{2022}]{2022ApJ...934...24C}
{Chen} J.~L.,  et~al., 2022, \mn@doi [\apj] {10.3847/1538-4357/ac75d1}, \href {https://ui.adsabs.harvard.edu/abs/2022ApJ...934...24C} {934, 24}

\bibitem[\protect\citeauthoryear{{Cordes} \& {Lazio}}{{Cordes} \& {Lazio}}{2002}]{ne2001}
{Cordes} J.~M.,  {Lazio} T.~J.~W.,  2002, arXiv e-prints, \href {https://ui.adsabs.harvard.edu/abs/2002astro.ph..7156C} {pp astro--ph/0207156}

\bibitem[\protect\citeauthoryear{{Cordes} \& {Shannon}}{{Cordes} \& {Shannon}}{2008}]{CS08}
{Cordes} J.~M.,  {Shannon} R.~M.,  2008, \mn@doi [\apj] {10.1086/589425}, \href {http://adsabs.harvard.edu/abs/2008ApJ...682.1152C} {682, 1152}

\bibitem[\protect\citeauthoryear{{Cordes} et~al.,}{{Cordes} et~al.}{2006}]{2006ApJ...637..446C}
{Cordes} J.~M.,  et~al., 2006, \mn@doi [\apj] {10.1086/498335}, \href {https://ui.adsabs.harvard.edu/abs/2006ApJ...637..446C} {637, 446}

\bibitem[\protect\citeauthoryear{{Cui}, {Boyles}, {McLaughlin}  \& {Palliyaguru}}{{Cui} et~al.}{2017}]{2017ApJ...840....5C}
{Cui} B.-Y.,  {Boyles} J.,  {McLaughlin} M.~A.,   {Palliyaguru} N.,  2017, \mn@doi [\apj] {10.3847/1538-4357/aa6aa9}, \href {https://ui.adsabs.harvard.edu/abs/2017ApJ...840....5C} {840, 5}

\bibitem[\protect\citeauthoryear{{Deneva} et~al.,}{{Deneva} et~al.}{2009}]{Deneva09}
{Deneva} J.~S.,  et~al., 2009, \mn@doi [\apj] {10.1088/0004-637X/703/2/2259}, \href {https://ui.adsabs.harvard.edu/abs/2009ApJ...703.2259D} {703, 2259}

\bibitem[\protect\citeauthoryear{{Deneva}, {Stovall}, {McLaughlin}, {Bates}, {Freire}, {Martinez}, {Jenet}  \& {Bagchi}}{{Deneva} et~al.}{2013}]{2013ApJ...775...51D}
{Deneva} J.~S.,  {Stovall} K.,  {McLaughlin} M.~A.,  {Bates} S.~D.,  {Freire} P.~C.~C.,  {Martinez} J.~G.,  {Jenet} F.,   {Bagchi} M.,  2013, \mn@doi [\apj] {10.1088/0004-637X/775/1/51}, \href {https://ui.adsabs.harvard.edu/abs/2013ApJ...775...51D} {775, 51}

\bibitem[\protect\citeauthoryear{{Deneva} et~al.,}{{Deneva} et~al.}{2016}]{Deneva16}
{Deneva} J.~S.,  et~al., 2016, \mn@doi [\apj] {10.3847/0004-637X/821/1/10}, \href {https://ui.adsabs.harvard.edu/abs/2016ApJ...821...10D} {821, 10}

\bibitem[\protect\citeauthoryear{{Dewey}, {Stokes}, {Segelstein}, {Taylor}  \& {Weisberg}}{{Dewey} et~al.}{1984}]{1984bens.work..234D}
{Dewey} R.,  {Stokes} G.,  {Segelstein} D.,  {Taylor} J.,   {Weisberg} J.,  1984, in {Reynolds} S.~P.,  {Stinebring} D.~R.,  eds, Birth and Evolution of Neutron Stars: Issues Raised by Millisecond Pulsars. p.~234

\bibitem[\protect\citeauthoryear{{Dong} et~al.,}{{Dong} et~al.}{2023}]{2023MNRAS.524.5132D}
{Dong} F.~A.,  et~al., 2023, \mn@doi [\mnras] {10.1093/mnras/stad2012}, \href {https://ui.adsabs.harvard.edu/abs/2023MNRAS.524.5132D} {524, 5132}

\bibitem[\protect\citeauthoryear{{Eatough}, {Keane}  \& {Lyne}}{{Eatough} et~al.}{2009}]{2009MNRAS.395..410E}
{Eatough} R.~P.,  {Keane} E.~F.,   {Lyne} A.~G.,  2009, \mn@doi [\mnras] {10.1111/j.1365-2966.2009.14524.x}, \href {https://ui.adsabs.harvard.edu/abs/2009MNRAS.395..410E} {395, 410}

\bibitem[\protect\citeauthoryear{{Edwards}, {Bailes}, {van Straten}  \& {Britton}}{{Edwards} et~al.}{2001}]{Edwards}
{Edwards} R.~T.,  {Bailes} M.,  {van Straten} W.,   {Britton} M.~C.,  2001, \mn@doi [\mnras] {10.1046/j.1365-8711.2001.04637.x}, \href {http://adsabs.harvard.edu/abs/2001MNRAS.326..358E} {326, 358}

\bibitem[\protect\citeauthoryear{{Faucher-Gigu{\`e}re} \& {Kaspi}}{{Faucher-Gigu{\`e}re} \& {Kaspi}}{2006}]{2006ApJ...643..332F}
{Faucher-Gigu{\`e}re} C.-A.,  {Kaspi} V.~M.,  2006, \mn@doi [\apj] {10.1086/501516}, \href {https://ui.adsabs.harvard.edu/abs/2006ApJ...643..332F} {643, 332}

\bibitem[\protect\citeauthoryear{{Good} et~al.,}{{Good} et~al.}{2021}]{2021ApJ...922...43G}
{Good} D.~C.,  et~al., 2021, \mn@doi [\apj] {10.3847/1538-4357/ac1da6}, \href {https://ui.adsabs.harvard.edu/abs/2021ApJ...922...43G} {922, 43}

\bibitem[\protect\citeauthoryear{Graber, Ronchi, Pardo-Araujo  \& Rea}{Graber et~al.}{2024}]{grp+2024}
Graber V.,  Ronchi M.,  Pardo-Araujo C.,   Rea N.,  2024, \mn@doi [The Astrophysical Journal] {10.3847/1538-4357/ad3e78}, 968, 16

\bibitem[\protect\citeauthoryear{{Han} et~al.,}{{Han} et~al.}{2021}]{2021RAA....21..107H}
{Han} J.~L.,  et~al., 2021, \mn@doi [Research in Astronomy and Astrophysics] {10.1088/1674-4527/21/5/107}, \href {https://ui.adsabs.harvard.edu/abs/2021RAA....21..107H} {21, 107}

\bibitem[\protect\citeauthoryear{{Han} et~al.,}{{Han} et~al.}{2025}]{2025RAA....25a4001H}
{Han} J.~L.,  et~al., 2025, \mn@doi [Research in Astronomy and Astrophysics] {10.1088/1674-4527/ada3b7}, \href {https://ui.adsabs.harvard.edu/abs/2025RAA....25a4001H} {25, 014001}

\bibitem[\protect\citeauthoryear{{Hessels}, {Ransom}, {Kaspi}, {Roberts}, {Champion}  \& {Stappers}}{{Hessels} et~al.}{2008}]{Hessels_2008GBT}
{Hessels} J.~W.~T.,  {Ransom} S.~M.,  {Kaspi} V.~M.,  {Roberts} M.~S.~E.,  {Champion} D.~J.,   {Stappers} B.~W.,  2008, in {Bassa} C.,  {Wang} Z.,  {Cumming} A.,   {Kaspi} V.~M.,  eds,  American Institute of Physics Conference Series Vol. 983, 40 Years of Pulsars: Millisecond Pulsars, Magnetars and More. pp 613--615 (\mn@eprint {arXiv} {0710.1745}), \mn@doi{10.1063/1.2900310}

\bibitem[\protect\citeauthoryear{{Jacoby}, {Bailes}, {Ord}, {Edwards}  \& {Kulkarni}}{{Jacoby} et~al.}{2009}]{Jacoby}
{Jacoby} B.~A.,  {Bailes} M.,  {Ord} S.~M.,  {Edwards} R.~T.,   {Kulkarni} S.~R.,  2009, \mn@doi [\apj] {10.1088/0004-637X/699/2/2009}, \href {http://adsabs.harvard.edu/abs/2009ApJ...699.2009J} {699, 2009}

\bibitem[\protect\citeauthoryear{{Johnston} \& {Karastergiou}}{{Johnston} \& {Karastergiou}}{2019}]{2019MNRAS.485..640J}
{Johnston} S.,  {Karastergiou} A.,  2019, \mn@doi [\mnras] {10.1093/mnras/stz400}, \href {https://ui.adsabs.harvard.edu/abs/2019MNRAS.485..640J} {485, 640}

\bibitem[\protect\citeauthoryear{{Karako-Argaman} et~al.,}{{Karako-Argaman} et~al.}{2015}]{GBNCC2015KA}
{Karako-Argaman} C.,  et~al., 2015, \mn@doi [\apj] {10.1088/0004-637X/809/1/67}, \href {https://ui.adsabs.harvard.edu/abs/2015ApJ...809...67K} {809, 67}

\bibitem[\protect\citeauthoryear{{Karuppusamy}, {Stappers}  \& {van Straten}}{{Karuppusamy} et~al.}{2010}]{2010A&A...515A..36K}
{Karuppusamy} R.,  {Stappers} B.~W.,   {van Straten} W.,  2010, \mn@doi [\aap] {10.1051/0004-6361/200913729}, \href {https://ui.adsabs.harvard.edu/abs/2010A&A...515A..36K} {515, A36}

\bibitem[\protect\citeauthoryear{{Keane}}{{Keane}}{2016}]{2016MNRAS.459.1360K}
{Keane} E.~F.,  2016, \mn@doi [\mnras] {10.1093/mnras/stw767}, \href {https://ui.adsabs.harvard.edu/abs/2016MNRAS.459.1360K} {459, 1360}

\bibitem[\protect\citeauthoryear{{Keane} \& {Kramer}}{{Keane} \& {Kramer}}{2008}]{2008MNRAS.391.2009K}
{Keane} E.~F.,  {Kramer} M.,  2008, \mn@doi [\mnras] {10.1111/j.1365-2966.2008.14045.x}, \href {https://ui.adsabs.harvard.edu/abs/2008MNRAS.391.2009K} {391, 2009}

\bibitem[\protect\citeauthoryear{{Keane} \& {McLaughlin}}{{Keane} \& {McLaughlin}}{2011}]{2011BASI...39..333K}
{Keane} E.~F.,  {McLaughlin} M.~A.,  2011, \mn@doi [Bulletin of the Astronomical Society of India] {10.48550/arXiv.1109.6896}, \href {https://ui.adsabs.harvard.edu/abs/2011BASI...39..333K} {39, 333}

\bibitem[\protect\citeauthoryear{{Keane}, {Ludovici}, {Eatough}, {Kramer}, {Lyne}, {McLaughlin}  \& {Stappers}}{{Keane} et~al.}{2010}]{2010MNRAS.401.1057K}
{Keane} E.~F.,  {Ludovici} D.~A.,  {Eatough} R.~P.,  {Kramer} M.,  {Lyne} A.~G.,  {McLaughlin} M.~A.,   {Stappers} B.~W.,  2010, \mn@doi [\mnras] {10.1111/j.1365-2966.2009.15693.x}, \href {https://ui.adsabs.harvard.edu/abs/2010MNRAS.401.1057K} {401, 1057}

\bibitem[\protect\citeauthoryear{Keane, Kramer, Lyne, Stappers  \& McLaughlin}{Keane et~al.}{2011}]{Keane2011}
Keane E.~F.,  Kramer M.,  Lyne A.~G.,  Stappers B.~W.,   McLaughlin M.~A.,  2011, \mn@doi [MNRAS] {10.1111/j.1365-2966.2011.18917.x}, 415, 3065

\bibitem[\protect\citeauthoryear{{Keane} et~al.,}{{Keane} et~al.}{2018}]{2018MNRAS.473..116K}
{Keane} E.~F.,  et~al., 2018, \mn@doi [\mnras] {10.1093/mnras/stx2126}, \href {https://ui.adsabs.harvard.edu/abs/2018MNRAS.473..116K} {473, 116}

\bibitem[\protect\citeauthoryear{{Keane} et~al.,}{{Keane} et~al.}{2025}]{2025OJAp....854256K}
{Keane} E.~F.,  et~al., 2025, \mn@doi [The Open Journal of Astrophysics] {10.33232/001c.154256}, \href {https://ui.adsabs.harvard.edu/abs/2025OJAp....854256K} {8, 54256}

\bibitem[\protect\citeauthoryear{{Keith} et~al.,}{{Keith} et~al.}{2010}]{HTRUmid}
{Keith} M.~J.,  et~al., 2010, \mn@doi [\mnras] {10.1111/j.1365-2966.2010.17325.x}, \href {http://adsabs.harvard.edu/abs/2010MNRAS.409..619K} {409, 619}

\bibitem[\protect\citeauthoryear{{Kramer}, {Karastergiou}, {Gupta}, {Johnston}, {Bhat}  \& {Lyne}}{{Kramer} et~al.}{2003}]{2003A&A...407..655K}
{Kramer} M.,  {Karastergiou} A.,  {Gupta} Y.,  {Johnston} S.,  {Bhat} N.~D.~R.,   {Lyne} A.~G.,  2003, \mn@doi [\aap] {10.1051/0004-6361:20030842}, \href {https://ui.adsabs.harvard.edu/abs/2003A&A...407..655K} {407, 655}

\bibitem[\protect\citeauthoryear{{Lazarus} et~al.,}{{Lazarus} et~al.}{2015}]{2015ApJ...812...81L}
{Lazarus} P.,  et~al., 2015, \mn@doi [\apj] {10.1088/0004-637X/812/1/81}, \href {https://ui.adsabs.harvard.edu/abs/2015ApJ...812...81L} {812, 81}

\bibitem[\protect\citeauthoryear{{Li}}{{Li}}{2006}]{Li2006}
{Li} X.-D.,  2006, \mn@doi [\apjl] {10.1086/506962}, \href {https://ui.adsabs.harvard.edu/abs/2006ApJ...646L.139L} {646, L139}

\bibitem[\protect\citeauthoryear{{Logvinenko}, {Tyul'bashev}  \& {Malofeev}}{{Logvinenko} et~al.}{2020}]{2020BLPI...47..390L}
{Logvinenko} S.~V.,  {Tyul'bashev} S.~A.,   {Malofeev} V.~M.,  2020, \mn@doi [Bulletin of the Lebedev Physics Institute] {10.3103/S1068335620120179}, \href {https://ui.adsabs.harvard.edu/abs/2020BLPI...47..390L} {47, 390}

\bibitem[\protect\citeauthoryear{{Lorimer} \& {Kramer}}{{Lorimer} \& {Kramer}}{2004}]{2004hpa..book.....L}
{Lorimer} D.~R.,  {Kramer} M.,  2004, {Handbook of Pulsar Astronomy}.
 Vol. 4

\bibitem[\protect\citeauthoryear{Lorimer et~al.,}{Lorimer et~al.}{2006}]{Lorimer2006}
Lorimer D.~R.,  et~al., 2006, \mn@doi [MNRAS] {10.1111/j.1365-2966.2006.10887.x}, 372, 777~(LFL06)

\bibitem[\protect\citeauthoryear{{Lynch} et~al.,}{{Lynch} et~al.}{2018}]{2018ApJ...859...93L}
{Lynch} R.~S.,  et~al., 2018, \mn@doi [\apj] {10.3847/1538-4357/aabf8a}, \href {https://ui.adsabs.harvard.edu/abs/2018ApJ...859...93L} {859, 93}

\bibitem[\protect\citeauthoryear{{Lyon}, {Stappers}, {Cooper}, {Brooke}  \& {Knowles}}{{Lyon} et~al.}{2016}]{2016MNRAS.459.1104L}
{Lyon} R.~J.,  {Stappers} B.~W.,  {Cooper} S.,  {Brooke} J.~M.,   {Knowles} J.~D.,  2016, \mn@doi [\mnras] {10.1093/mnras/stw656}, \href {https://ui.adsabs.harvard.edu/abs/2016MNRAS.459.1104L} {459, 1104}

\bibitem[\protect\citeauthoryear{{Ma} et~al.,}{{Ma} et~al.}{2025}]{2025A&A...698A.306M}
{Ma} X.,  et~al., 2025, \mn@doi [\aap] {10.1051/0004-6361/202452685}, \href {https://ui.adsabs.harvard.edu/abs/2025A&A...698A.306M} {698, A306}

\bibitem[\protect\citeauthoryear{{Manchester} et~al.,}{{Manchester} et~al.}{2001}]{pmsurv}
{Manchester} R.~N.,  et~al., 2001, \mn@doi [\mnras] {10.1046/j.1365-8711.2001.04751.x}, \href {http://adsabs.harvard.edu/abs/2001MNRAS.328...17M} {328, 17}

\bibitem[\protect\citeauthoryear{{McKenna}, {Keane}, {Gallagher}  \& {McCauley}}{{McKenna} et~al.}{2024}]{2024MNRAS.527.4397M}
{McKenna} D.~J.,  {Keane} E.~F.,  {Gallagher} P.~T.,   {McCauley} J.,  2024, \mn@doi [\mnras] {10.1093/mnras/stad2900}, \href {https://ui.adsabs.harvard.edu/abs/2024MNRAS.527.4397M} {527, 4397}

\bibitem[\protect\citeauthoryear{{McLaughlin} \& {Cordes}}{{McLaughlin} \& {Cordes}}{2003}]{McLaughlin_Cordes_2003}
{McLaughlin} M.~A.,  {Cordes} J.~M.,  2003, \mn@doi [\apj] {10.1086/378232}, \href {https://ui.adsabs.harvard.edu/abs/2003ApJ...596..982M} {596, 982}

\bibitem[\protect\citeauthoryear{McLaughlin et~al.,}{McLaughlin et~al.}{2006}]{McLaughlin2006}
McLaughlin M.~A.,  et~al., 2006, \mn@doi [Nature] {10.1038/nature04440}, 439, 817

\bibitem[\protect\citeauthoryear{{Mcsweeney}, {Moseley}, {Hurley-Walker}, {Grover}, {Horv{\'a}th}, {Galvin}, {Meyers}  \& {Tan}}{{Mcsweeney} et~al.}{2025}]{2025ApJ...981..143M}
{Mcsweeney} S.~J.,  {Moseley} J.,  {Hurley-Walker} N.,  {Grover} G.,  {Horv{\'a}th} C.,  {Galvin} T.~J.,  {Meyers} B.~W.,   {Tan} C.~M.,  2025, \mn@doi [\apj] {10.3847/1538-4357/adb27f}, \href {https://ui.adsabs.harvard.edu/abs/2025ApJ...981..143M} {981, 143}

\bibitem[\protect\citeauthoryear{{Michilli} et~al.,}{{Michilli} et~al.}{2020}]{2020MNRAS.491..725M}
{Michilli} D.,  et~al., 2020, \mn@doi [\mnras] {10.1093/mnras/stz2997}, \href {https://ui.adsabs.harvard.edu/abs/2020MNRAS.491..725M} {491, 725}

\bibitem[\protect\citeauthoryear{{Mickaliger}, {McEwen}, {McLaughlin}  \& {Lorimer}}{{Mickaliger} et~al.}{2018}]{Mitch2018}
{Mickaliger} M.~B.,  {McEwen} A.~E.,  {McLaughlin} M.~A.,   {Lorimer} D.~R.,  2018, \mn@doi [\mnras] {10.1093/mnras/sty1785}, \href {https://ui.adsabs.harvard.edu/abs/2018MNRAS.479.5413M} {479, 5413}

\bibitem[\protect\citeauthoryear{{Morello} et~al.,}{{Morello} et~al.}{2019}]{2019MNRAS.483.3673M}
{Morello} V.,  et~al., 2019, \mn@doi [\mnras] {10.1093/mnras/sty3328}, \href {https://ui.adsabs.harvard.edu/abs/2019MNRAS.483.3673M} {483, 3673}

\bibitem[\protect\citeauthoryear{{Morello}, {Barr}, {Stappers}, {Keane}  \& {Lyne}}{{Morello} et~al.}{2020}]{2020MNRAS.497.4654M}
{Morello} V.,  {Barr} E.~D.,  {Stappers} B.~W.,  {Keane} E.~F.,   {Lyne} A.~G.,  2020, \mn@doi [\mnras] {10.1093/mnras/staa2291}, \href {https://ui.adsabs.harvard.edu/abs/2020MNRAS.497.4654M} {497, 4654}

\bibitem[\protect\citeauthoryear{{Murphy} \& {Kaplan}}{{Murphy} \& {Kaplan}}{2025}]{2025arXiv251110785M}
{Murphy} T.,  {Kaplan} D.~L.,  2025, \mn@doi [arXiv e-prints] {10.48550/arXiv.2511.10785}, \href {https://ui.adsabs.harvard.edu/abs/2025arXiv251110785M} {p. arXiv:2511.10785}

\bibitem[\protect\citeauthoryear{{Nice} et~al.,}{{Nice} et~al.}{2013}]{2013ApJ...772...50N}
{Nice} D.~J.,  et~al., 2013, \mn@doi [\apj] {10.1088/0004-637X/772/1/50}, \href {https://ui.adsabs.harvard.edu/abs/2013ApJ...772...50N} {772, 50}

\bibitem[\protect\citeauthoryear{{Ocker} \& {Cordes}}{{Ocker} \& {Cordes}}{2026}]{2026arXiv260211838O}
{Ocker} S.~K.,  {Cordes} J.~M.,  2026, \mn@doi [arXiv e-prints] {10.48550/arXiv.2602.11838}, \href {https://ui.adsabs.harvard.edu/abs/2026arXiv260211838O} {p. arXiv:2602.11838}

\bibitem[\protect\citeauthoryear{{Padmanabh} et~al.,}{{Padmanabh} et~al.}{2023}]{2023MNRAS.524.1291P}
{Padmanabh} P.~V.,  et~al., 2023, \mn@doi [\mnras] {10.1093/mnras/stad1900}, \href {https://ui.adsabs.harvard.edu/abs/2023MNRAS.524.1291P} {524, 1291}

\bibitem[\protect\citeauthoryear{{Palliyaguru} et~al.,}{{Palliyaguru} et~al.}{2011}]{Nipuni2011}
{Palliyaguru} N.~T.,  et~al., 2011, \mn@doi [\mnras] {10.1111/j.1365-2966.2011.19388.x}, \href {https://ui.adsabs.harvard.edu/abs/2011MNRAS.417.1871P} {417, 1871}

\bibitem[\protect\citeauthoryear{{Parent}}{{Parent}}{2022}]{eparent_thesis}
{Parent} E.,  2022, PhD thesis, McGill University

\bibitem[\protect\citeauthoryear{{Parent} et~al.,}{{Parent} et~al.}{2022}]{2022ApJ...924..135P}
{Parent} E.,  et~al., 2022, \mn@doi [\apj] {10.3847/1538-4357/ac375d}, \href {https://ui.adsabs.harvard.edu/abs/2022ApJ...924..135P} {924, 135}

\bibitem[\protect\citeauthoryear{Patel et~al.,}{Patel et~al.}{2018}]{Patel2018}
Patel C.,  et~al., 2018, \mn@doi [The Astrophysical Journal] {10.3847/1538-4357/aaee65}, 869, 181

\bibitem[\protect\citeauthoryear{Philippov \& Kramer}{Philippov \& Kramer}{2022}]{pk22}
Philippov S.,  Kramer M.,  2022, \mn@doi [Annual Review of Astronomy and Astrophysics] {10.1146/annurev-astro-052920-112338}, 60, 495

\bibitem[\protect\citeauthoryear{{Phinney} \& {Blandford}}{{Phinney} \& {Blandford}}{1981}]{1981MNRAS.194..137P}
{Phinney} E.~S.,  {Blandford} R.~D.,  1981, \mn@doi [\mnras] {10.1093/mnras/194.1.137}, \href {https://ui.adsabs.harvard.edu/abs/1981MNRAS.194..137P} {194, 137}

\bibitem[\protect\citeauthoryear{Qiu, Bannister, Shannon, Murphy, Bhandari, Agarwal, Lorimer  \& Bunton}{Qiu et~al.}{2019}]{Qiu_2019}
Qiu H.,  Bannister K.~W.,  Shannon R.~M.,  Murphy T.,  Bhandari S.,  Agarwal D.,  Lorimer D.~R.,   Bunton J.~D.,  2019, \mn@doi [MNRAS] {10.1093/mnras/stz748}, 486, 166

\bibitem[\protect\citeauthoryear{{Rane} \& {Loeb}}{{Rane} \& {Loeb}}{2016}]{2016arXiv160806952R}
{Rane} A.,  {Loeb} A.,  2016, arXiv e-prints, \href {https://ui.adsabs.harvard.edu/abs/2016arXiv160806952R} {p. arXiv:1608.06952}

\bibitem[\protect\citeauthoryear{{Ravi}, {Hallinan}  \& {Deep Synoptic Array Team}}{{Ravi} et~al.}{2021}]{2021AAS...23731604R}
{Ravi} V.,  {Hallinan} G.,   {Deep Synoptic Array Team} 2021, in American Astronomical Society Meeting Abstracts \#237. p. 316.04

\bibitem[\protect\citeauthoryear{{Rozwadowska}, {Vissani}  \& {Cappellaro}}{{Rozwadowska} et~al.}{2021}]{2021NewA...8301498R}
{Rozwadowska} K.,  {Vissani} F.,   {Cappellaro} E.,  2021, \mn@doi [\na] {10.1016/j.newast.2020.101498}, \href {https://ui.adsabs.harvard.edu/abs/2021NewA...8301498R} {83, 101498}

\bibitem[\protect\citeauthoryear{{Ruderman} \& {Sutherland}}{{Ruderman} \& {Sutherland}}{1975}]{rs75}
{Ruderman} M.,  {Sutherland} P.~G.,  1975, \apj, 196, 51

\bibitem[\protect\citeauthoryear{{Samodurov}, {Tyul'bashev}, {Toropov}, {Dolgushev}, {Oreshko}  \& {Logvinenko}}{{Samodurov} et~al.}{2023}]{2023ARep...67..590S}
{Samodurov} V.~A.,  {Tyul'bashev} S.~A.,  {Toropov} M.~O.,  {Dolgushev} A.~V.,  {Oreshko} V.~V.,   {Logvinenko} S.~V.,  2023, \mn@doi [Astronomy Reports] {10.1134/S1063772923070077}, \href {https://ui.adsabs.harvard.edu/abs/2023ARep...67..590S} {67, 590}

\bibitem[\protect\citeauthoryear{{Schechter}}{{Schechter}}{1976}]{1976ApJ...203..297S}
{Schechter} P.,  1976, \mn@doi [\apj] {10.1086/154079}, \href {https://ui.adsabs.harvard.edu/abs/1976ApJ...203..297S} {203, 297}

\bibitem[\protect\citeauthoryear{{Sengar} et~al.,}{{Sengar} et~al.}{2023}]{2023MNRAS.522.1071S}
{Sengar} R.,  et~al., 2023, \mn@doi [\mnras] {10.1093/mnras/stad508}, \href {https://ui.adsabs.harvard.edu/abs/2023MNRAS.522.1071S} {522, 1071}

\bibitem[\protect\citeauthoryear{{Shapiro-Albert}, {McLaughlin}  \& {Keane}}{{Shapiro-Albert} et~al.}{2018}]{Brent2018}
{Shapiro-Albert} B.~J.,  {McLaughlin} M.~A.,   {Keane} E.~F.,  2018, \mn@doi [\apj] {10.3847/1538-4357/aae2b2}, \href {https://ui.adsabs.harvard.edu/abs/2018ApJ...866..152S} {866, 152}

\bibitem[\protect\citeauthoryear{{Shitov}, {Kuzmin}, {Dumskii}  \& {Losovsky}}{{Shitov} et~al.}{2009}]{2009ARep...53..561S}
{Shitov} Y.~P.,  {Kuzmin} A.~D.,  {Dumskii} D.~V.,   {Losovsky} B.~Y.,  2009, \mn@doi [Astronomy Reports] {10.1134/S1063772909060080}, \href {https://ui.adsabs.harvard.edu/abs/2009ARep...53..561S} {53, 561}

\bibitem[\protect\citeauthoryear{{Staelin}}{{Staelin}}{1969}]{1969IEEEP..57..724S}
{Staelin} D.~H.,  1969, \mn@doi [IEEE Proceedings] {10.1109/PROC.1969.7051}, \href {https://ui.adsabs.harvard.edu/abs/1969IEEEP..57..724S} {57, 724}

\bibitem[\protect\citeauthoryear{{Tauris} \& {Manchester}}{{Tauris} \& {Manchester}}{1998}]{1998MNRAS.298..625T}
{Tauris} T.~M.,  {Manchester} R.~N.,  1998, \mn@doi [\mnras] {10.1046/j.1365-8711.1998.01369.x}, \href {https://ui.adsabs.harvard.edu/abs/1998MNRAS.298..625T} {298, 625}

\bibitem[\protect\citeauthoryear{{Tian} et~al.,}{{Tian} et~al.}{2025}]{2025MNRAS.544.1843T}
{Tian} J.,  et~al., 2025, \mn@doi [\mnras] {10.1093/mnras/staf1827}, \href {https://ui.adsabs.harvard.edu/abs/2025MNRAS.544.1843T} {544, 1843}

\bibitem[\protect\citeauthoryear{{Turner} et~al.,}{{Turner} et~al.}{2024}]{2024MNRAS.531.3579T}
{Turner} J.~D.,  et~al., 2024, \mn@doi [\mnras] {10.1093/mnras/stae1300}, \href {https://ui.adsabs.harvard.edu/abs/2024MNRAS.531.3579T} {531, 3579}

\bibitem[\protect\citeauthoryear{{Turner} et~al.,}{{Turner} et~al.}{2025}]{2025MNRAS.537.1070T}
{Turner} J.~D.,  et~al., 2025, \mn@doi [\mnras] {10.1093/mnras/staf098}, \href {https://ui.adsabs.harvard.edu/abs/2025MNRAS.537.1070T} {537, 1070}

\bibitem[\protect\citeauthoryear{{Tyul'bashev} et~al.,}{{Tyul'bashev} et~al.}{2018a}]{2018ARep...62...63T}
{Tyul'bashev} S.~A.,  et~al., 2018a, \mn@doi [Astronomy Reports] {10.1134/S1063772918010079}, \href {https://ui.adsabs.harvard.edu/abs/2018ARep...62...63T} {62, 63}

\bibitem[\protect\citeauthoryear{{Tyul'bashev}, {Tyul'bashev}  \& {Malofeev}}{{Tyul'bashev} et~al.}{2018b}]{2018AA...618A..70T}
{Tyul'bashev} S.~A.,  {Tyul'bashev} V.~S.,   {Malofeev} V.~M.,  2018b, \mn@doi [\aap] {10.1051/0004-6361/201833102}, \href {https://ui.adsabs.harvard.edu/abs/2018A&A...618A..70T} {618, A70}

\bibitem[\protect\citeauthoryear{{Tyul'bashev}, {Kitaeva}, {Logvinenko}  \& {Tyul'basheva}}{{Tyul'bashev} et~al.}{2021}]{2021ARep...65.1246T}
{Tyul'bashev} S.~A.,  {Kitaeva} M.~A.,  {Logvinenko} S.~V.,   {Tyul'basheva} G.~E.,  2021, \mn@doi [Astronomy Reports] {10.1134/S1063772921120088}, \href {https://ui.adsabs.harvard.edu/abs/2021ARep...65.1246T} {65, 1246}

\bibitem[\protect\citeauthoryear{{Tyul'bashev}, {Kitaeva}, {Pervoukhin}, {Tyul'basheva}, {Brylyakova}, {Chernosov}  \& {Ovchinnikov}}{{Tyul'bashev} et~al.}{2024}]{2024AA...689A...1T}
{Tyul'bashev} S.~A.,  {Kitaeva} M.~A.,  {Pervoukhin} D.~V.,  {Tyul'basheva} G.~E.,  {Brylyakova} E.~A.,  {Chernosov} A.~V.,   {Ovchinnikov} I.~L.,  2024, \mn@doi [\aap] {10.1051/0004-6361/202449762}, \href {https://ui.adsabs.harvard.edu/abs/2024A&A...689A...1T} {689, A1}

\bibitem[\protect\citeauthoryear{{Vivekanand} \& {Narayan}}{{Vivekanand} \& {Narayan}}{1981}]{1981JApA....2..315V}
{Vivekanand} M.,  {Narayan} R.,  1981, \mn@doi [Journal of Astrophysics and Astronomy] {10.1007/BF02714556}, \href {https://ui.adsabs.harvard.edu/abs/1981JApA....2..315V} {2, 315}

\bibitem[\protect\citeauthoryear{{Walter}, {Hallinan}, {Ravi}, {McLaughlin}, {Jameson}  \& {Kapp}}{{Walter} et~al.}{2025}]{2025AAS...24614305W}
{Walter} F.,  {Hallinan} G.,  {Ravi} V.,  {McLaughlin} M.,  {Jameson} K.,   {Kapp} F.,  2025, in 246th Meeting of the American Astronomical Society. p. 143.05

\bibitem[\protect\citeauthoryear{{Weltevrede}, {Stappers}, {Rankin}  \& {Wright}}{{Weltevrede} et~al.}{2006}]{2006ApJ...645L.149W}
{Weltevrede} P.,  {Stappers} B.~W.,  {Rankin} J.~M.,   {Wright} G.~A.~E.,  2006, \mn@doi [\apjl] {10.1086/506346}, \href {https://ui.adsabs.harvard.edu/abs/2006ApJ...645L.149W} {645, L149}

\bibitem[\protect\citeauthoryear{{Yao}, {Manchester}  \& {Wang}}{{Yao} et~al.}{2017}]{ymw16}
{Yao} J.~M.,  {Manchester} R.~N.,   {Wang} N.,  2017, \mn@doi [\apj] {10.3847/1538-4357/835/1/29}, \href {https://ui.adsabs.harvard.edu/abs/2017ApJ...835...29Y} {835, 29}

\bibitem[\protect\citeauthoryear{{Zhou} et~al.,}{{Zhou} et~al.}{2023}]{2023RAA....23j4001Z}
{Zhou} D.~J.,  et~al., 2023, \mn@doi [Research in Astronomy and Astrophysics] {10.1088/1674-4527/accc76}, \href {https://ui.adsabs.harvard.edu/abs/2023RAA....23j4001Z} {23, 104001}

\makeatother
\end{thebibliography}

\appendix

\section{The \textsc{RRATalog}}

Table A1 provides the primary observational parameters for the sample of 335 currently known RRATs. From left to right, we list the RRAT name, dispersion measure (DM), spin period ($P$), sky location in Galactic coordinates ($l, b$), and the observed burst rate (${\cal B}$). For the subset of 37 RRATs with established timing solutions, Table A2 additionally provides the right ascension and declination, the measured period, epoch of period and period derivative ($\dot{P}$), and several derived physical quantities. To compute these derived quantities, we assume \citep[see, e.g.,][]{2004hpa..book.....L} a standard dipole model for a 1.4~M$_\odot$ neutron star of radius 10~km. Under these assumptions, the surface magnetic field
\begin{equation}
B_{\text{s}} = 1.0 \times 10^{12} \left(\frac{P}{\rm s}\right)^{1/2} \left(\frac{\dot{P}}{10^{-15}\,{\rm s~s}^{-1}}\right)^{1/2} \,{\rm G},
\end{equation}
the magnetic field at the light cylinder 
\begin{equation}
B_{\text{LC}} = 4.8 \left(\frac{P}{\rm s}\right)^{-5/2} \left(\frac{\dot{P}}{10^{-15}\,{\rm s~s}^{-1}}\right)^{1/2} \,{\rm G},
\end{equation}
the spin-down energy loss rate 
\begin{equation}
\dot{E} =  3.95 \times 10^{31}  \left(\frac{P}{\rm s}\right)^{-3} \left(\frac{\dot{P}}{10^{-15}\,{\rm s~s}^{-1}}\right) \, {\rm erg~s^{-1}}
\end{equation}
and the characteristic age 
\begin{equation}
\tau = 15.8 \left(\frac{P}{\rm s}\right) \left(\frac{\dot{P}}{10^{-15}\,{\rm s~s}^{-1}}\right)^{-1} \,{\rm My}.
\end{equation}

\begin{table*}
    \centering
    \small
\begin{tabular}{lllrrrrrl}
\toprule
      RRAT &   DM  &         $P$ & \multicolumn{1}{c}{$l$} & \multicolumn{1}{c}{$b$} & ${\cal B}$ & ${S}_{1400}$  & ${W}_{1400}$ & Reference \\
     &  (cm$^{-3}$~pc) &         (s) & \multicolumn{1}{c}{(\degr)} & \multicolumn{1}{c}{(\degr)} & (hr$^{-1}$) & (Jy) & (ms) & \\\midrule
  J0012+5431 & 131.3(7) & 3.025 & 117.23 & --7.91 & 2.7 & -- & -- & \citet{2023MNRAS.524.5132D} \\
{J0031--5726} & {6.75(3)} & {1.570} & {308.21} & {--59.48} & {60} & {--} & {--} & {\citet{2025ApJ...981..143M}} \\ 
{J0054+6650} & {14.560(2)} & {1.390} & {123.28} & {3.97} & {0.7} & {--} & {--} & {\citet{Hessels_2008GBT}} \\
  J0054+69 & 90.3(2) & -- & 123.20 & 6.56 & 300.0 & -- & -- & \citet{GBNCC2015KA} \\
{J0102+5356} & {55.6200(8)} & {0.354} & {124.65} & {--8.90} & {390.0} & {--} & {--} & {\citet{GBNCC2015KA}} \\
  J0121+53 & 91.38(3) & 2.725 & 127.38 & --9.13 & 2.4 & -- & -- & \citet{2021ApJ...922...43G} \\
  J0139+3336 & 21.23(1) & 1.248 & 134.38 & --28.17 & -- & -- & -- & \citet{2018ARep...62...63T} \\
  J0156+04 & 27.5 & 1.359 & 151.97 & --55.15 & 2 & -- & -- & \citet{Deneva16} \\
  J0201+7005 & 21.029(2) & 1.349 & 128.89 & 8.03 & 180 & -- & -- & \citet{GBNCC2015KA} \\
  J0219--06 & 8.46(7) & 1.879 & 171.68 & --60.46 & -- & -- & -- & \citet{2025MNRAS.537.1070T} \\
  J0249+52 & 27.5(1.5) & -- & 140.31 & --6.09 & 0.44 & -- & -- & \citet{2023ARep...67..590S} \\
  J0302+2252 & 18.9922(6) & 1.207 & 158.44 & --30.82 & -- & -- & -- & \citet{2020MNRAS.491..725M} \\
  J0305+40 & 24(2) & -- & 149.01 & --15.97 & -- & -- & -- & \citet{2018AA...618A..70T} \\
  J0313+36 & 20.8(1.5) & -- & 152.45 & --18.37 & -- & -- & -- & \citet{2024AA...689A...1T} \\
{J0317+1328} & {12.7452(10)} & {1.974} & {168.76} & {--36.04} & {--} & {--} & {--} & {\citet{2018ARep...62...63T}} \\
  J0332+79 & 16.67(2) & 2.056 & 130.31 & 18.68 & 30 & -- & -- & \citet{GBNCC2015KA} \\
  J0402--6542 & 31.5(2) & 3.034 & 278.94 & --41.40 & -- & -- & -- & \citet{2025MNRAS.537.1070T} \\
  J0408+28 & 2.5(1.0) & 2.920 & 166.90 & --16.84 & -- & -- & -- & \citet{2024AA...689A...1T} \\
  J0410--31 & 9.2(3) & 1.879 & 230.59 & --46.67 & 107 & -- & -- & \citet{HTRU-RRATs} \\
  J0440+35 & 2.6(1.0) & 2.230 & 166.63 & --7.46 & -- & -- & -- & \citet{2024AA...689A...1T} \\
  J0441--04 & 20.0 & -- & 200.95 & --30.83 & -- & -- & -- & \citet{GBNCC2015KA} \\
  J0447--04 & 29.83(4) & 2.188 & 202.07 & --29.66 & 103 & -- & -- & \citet{GBNCC2015KA} \\
  J0452+16 & 19(3) & -- & 183.27 & --16.79 & -- & -- & -- & \citet{2018AA...618A..70T} \\
  J0513--04 & 18.5 & -- & 205.23 & --23.81 & -- & -- & -- & \citet{GBNCC2015KA} \\
  J0534+34 & 24.5(1.5) & -- & 174.35 & 0.75 & -- & -- & -- & \citet{2018AA...618A..70T} \\
  J0544+20 & 56.9 & -- & 186.76 & --4.50 & 4 & -- & -- & \citet{Deneva16} \\
  J0545--03 & 67.2(4) & 1.074 & 208.09 & --16.21 & 77 & -- & -- & \citet{GBNCC2015KA} \\
  J0550+09 & 86.6 & 1.745 & 197.06 & --8.77 & 47 & -- & -- & \citet{Deneva16} \\
  J0601+38 & 20.9(1.5) & -- & 173.09 & 7.71 & -- & -- & -- & \citet{2024AA...689A...1T} \\
  J0614--03 & 17.9 & 0.136 & 211.89 & --9.69 & -- & -- & -- & \citet{GBNCC2015KA} \\
  J0621--55 & 22 & -- & 264.80 & --26.41 & -- & -- & -- & \citet{2018MNRAS.473..116K} \\
  J0623+1536 & 92.7 & 2.639 & 195.89 & 1.01 & -- & 0.032 & 14.1 & \citet{Patel2018} \\
  J0625+1254 & 101.9(6.1) & -- & 198.51 & 0.20 & -- & 0.036 & 7.1 & \citet{Patel2018} \\
  J0625+17 & 58(4) & 2.518 & 194.43 & 2.33 & -- & -- & -- & \citet{2018AA...618A..70T} \\
  J0627+16 & 113.0 & 2.180 & 195.79 & 2.12 & 23 & 0.15 & 0.3 & \citet{Deneva09} \\
  J0628+0909 & 88.3(2) & 1.241 & 202.19 & --0.85 & 141 & 0.085 & 10 & \citet{2006ApJ...637..446C} \\
  J0630+1933 & 47.2 & 1.249 & 193.13 & 4.27 & -- & -- & -- & \citet{Deneva16} \\
  J0630+23 & 12.4(1.0) & -- & 189.92 & 6.08 & -- & -- & -- & \citet{2024AA...689A...1T} \\
  J0637+0332 & 152(2) & -- & 208.22 & --1.44 & -- & -- & -- & \citet{2023RAA....23j4001Z} \\
  J0639+0828 & 290.1 & -- & 204.08 & 1.30 & -- & -- & -- & \citet{2025RAA....25a4001H} \\
  J0640+07 & 52(3) & -- & 204.83 & 1.14 & -- & -- & -- & \citet{2018ARep...62...63T} \\
  J0653--06 & 83.7 & 0.790 & 218.69 & --2.51 & 1.4 & -- & -- & \citet{2023MNRAS.524.5132D} \\
  J0657--46 & 148.4(4) & -- & 256.94 & --18.51 & -- & -- & -- & \citet{2025MNRAS.544.1843T} \\
  J0723--2050 & 130 & 0.712 & 234.99 & --2.73 & -- & -- & -- & \citet{2022MNRAS.512.1483B} \\
  J0736--6304 & 19.4 & 4.863 & 274.88 & --19.15 & 39.65 & 0.16 & 30 & \citet{BS10} \\
  J0741+17 & 44.3 & 1.730 & 202.77 & 18.44 & 3.7 & -- & -- & \citet{2023MNRAS.524.5132D} \\
{J0746+5514} & {10.318(7)} & {2.894} & {162.54} & {29.73} & {1.11} & {--} & {--} & {\citet{2024MNRAS.527.4397M}} \\
  J0803+34 & 34(2) & -- & 186.94 & 28.85 & -- & -- & -- & \citet{2018AA...618A..70T} \\
  J0812+8626 & 40.2(2) & -- & 126.73 & 28.31 & -- & -- & -- & \citet{2021ARep...65.1246T} \\
  J0837--24 & 142.8(5) & -- & 247.45 & 9.80 & 5 & 0.42 & 1 & \citet{HTRU-RRATs} \\
  J0845--36 & 29(2) & -- & 257.40 & 4.26 & 1.8 & 0.23 & 2 & \citet{Keane2011} \\
  J0847--4316 & 292.5(9) & 5.977 & 263.44 & 0.16 & 1.42 & 0.12 & 27 & \citet{McLaughlin2006} \\
  J0912--3851 & 71.5(7) & 1.526 & 263.16 & 6.58 & 32 & -- & 35.609 & \citet{HTRU-RRATs} \\
  J0917--4245 & 97.7(3) & 2.552 & 266.64 & 4.54 & -- & -- & -- & \citet{2025MNRAS.537.1070T} \\
  J0917--4420 & 45.8(1) & 2.581 & 267.83 & 3.49 & -- & -- & -- & \citet{2025MNRAS.544.1843T} \\
  J0923--31 & 72(20) & -- & 259.70 & 13.00 & 1.7 & 0.12 & 30 & \citet{BS10} \\
  J0930--1854 & 33 & -- & 250.76 & 23.02 & -- & -- & -- & \citet{2022MNRAS.512.1483B} \\
  J0933--4604 & 120.8(1) & 3.670 & 271.06 & 4.21 & -- & -- & -- & \citet{2025MNRAS.544.1843T} \\
  J0941+16 & 23(2) & -- & 216.58 & 44.87 & -- & -- & -- & \citet{2018AA...618A..70T} \\
  J0941--39 & 78.2(2.7) & 0.587 & 267.80 & 9.90 & -- & 0.58 & 105.6 & \citet{BS10} \\
  J0943--5305 & 174 & 1.734 & 276.84 & --0.03 & -- & -- & -- & \citet{2022MNRAS.512.1483B} \\
  J0957--06 & 26.95(2) & 1.724 & 244.76 & 36.20 & 180 & -- & -- & \citet{GBNCC2015KA} \\
\bottomrule
\end{tabular}
    \caption{Positions, dispersion measures (with uncertainties in the least significant digits when available), spin periods (rounded to three decimal places), galactic latitude and longitude, burst rates, flux densities, pulse widths, and references to the discovery paper for each RRAT in the \textsc{RRATalog}.}
\end{table*}
\addtocounter{table}{-1}
\begin{table*}
    \centering
    \small
\begin{tabular}{lllrrrrrl}
\toprule
      RRAT &   DM  &         $P$ & \multicolumn{1}{c}{$l$} & \multicolumn{1}{c}{$b$} & ${\cal B}$ & ${S}_{1400}$  & ${W}_{1400}$ & Reference \\
     &  (cm$^{-3}$~pc) &         (s) & \multicolumn{1}{c}{(\degr)} & \multicolumn{1}{c}{(\degr)} & (hr$^{-1}$) & (Jy) & (ms) & \\\midrule
{J1006+3015} & {18.085(4)} & {3.066} & {197.95} & {53.91} & {1.74} & {--} & {--} & {\citet{2018AA...618A..70T}} \\
  J1010+15 & 42 & -- & 221.26 & 50.98 & -- & -- & -- & \citet{2013ApJ...775...51D} \\
  J1014--48 & 87(7) & 1.509 & 278.14 & 6.33 & 16 & 0.14 & 21 & \citet{HTRU-RRATs} \\
  J1046--59 & 101.1(7) & -- & 287.62 & --0.21 & -- & -- & -- & \citet{2025MNRAS.544.1843T} \\
  J1048--5838 & 70.7(9) & 1.231 & 287.47 & 0.48 & 6.0 & 0.63 & 7 & \citet{2010MNRAS.401.1057K} \\
  J1059--01 & 18.7 & -- & 254.53 & 50.96 & -- & -- & -- & \citet{GBNCC2015KA} \\
  J1104+14 & 23.2(1.5) & -- & 234.88 & 61.90 & -- & -- & -- & \citet{2024AA...689A...1T} \\
  J1105+02 & 16.5(4) & 6.403 & 252.59 & 54.65 & 2.5 & -- & -- & \citet{2023MNRAS.524.5132D} \\
  J1108--5946 & 92.7(4) & 1.517 & 290.25 & 0.52 & -- & -- & -- & \citet{2025MNRAS.537.1070T} \\
  J1111--55 & 235(5) & -- & 288.79 & 5.09 & 0.4 & 0.08 & 16 & \citet{Keane2011} \\
  J1126--27 & 26.86(7) & 0.358 & 280.68 & 31.54 & 180 & -- & -- & \citet{GBNCC2015KA} \\
  J1129--53 & 77.0(2.5) & 1.063 & 290.80 & 7.41 & 36.2 & 0.32 & 19.1 & \citet{BS10} \\
  J1130+0921 & 21.0(9) & 4.797 & 252.22 & 63.98 & 2.9 & -- & -- & \citet{2023MNRAS.524.5132D} \\
{J1132+2513} & {24.2(1)} & {1.002} & {214.58} & {72.28} & {--} & {--} & {--} & {\citet{2018AA...618A..70T}} \\
  J1135--49 & 114(20) & -- & 290.53 & 11.62 & 1.3 & 0.12 & 9 & \citet{HTRU-RRATs} \\
  J1152--6056 & 381 & 2.449 & 295.85 & 1.12 & -- & -- & -- & \citet{2022MNRAS.512.1483B} \\
  J1153--21 & 34.8(1) & 2.343 & 285.19 & 39.55 & 150 & -- & -- & \citet{GBNCC2015KA} \\
  J1157+25 & 8.85(1.0) & -- & 216.83 & 77.89 & -- & -- & -- & \citet{2024AA...689A...1T} \\
  J1216--50 & 110(20) & 6.355 & 297.23 & 12.03 & 13 & 0.13 & 9 & \citet{HTRU-RRATs} \\
  J1226--3223 & 36.7 & 6.193 & 296.91 & 30.20 & 40.9 & 0.27 & 34 & \citet{BS10} \\
  J1243--0435 & 12.0(1) & 4.868 & 299.12 & 58.22 & -- & -- & -- & \citet{2025MNRAS.544.1843T} \\
  J1243--64 & 342(2) & -- & 302.09 & --1.53 & -- & -- & -- & \citet{2025MNRAS.537.1070T} \\
  J1252+53 & 20.7(3) & 0.220 & 122.75 & 63.43 & 0.09 & -- & -- & \citet{2021ApJ...922...43G} \\
  J1303--4713 & 82.6(1) & -- & 305.06 & 15.60 & -- & -- & -- & \citet{2025MNRAS.544.1843T} \\
  J1307--67 & 44(2) & 3.651 & 304.52 & --4.24 & 11 & 0.07 & 2 & \citet{Keane2011} \\
  J1308--61 & 224.5(2) & 3.955 & 305.04 & 1.52 & -- & -- & -- & \citet{2025MNRAS.537.1070T} \\
  J1311--59 & 152(5) & -- & 305.45 & 3.78 & 0.3 & 0.13 & 16 & \citet{Keane2011} \\
  J1317--5759 & 145.3(3) & 2.642 & 306.43 & 4.70 & 4.5 & 0.38 & 12 & \citet{McLaughlin2006} \\
  J1319--4536 & 40.41(8) & 1.871 & 308.11 & 16.99 & -- & -- & -- & \citet{2025MNRAS.537.1070T} \\
  J1332--03 & 27.1(2) & 1.106 & 322.25 & 57.91 & 51 & -- & -- & \citet{GBNCC2015KA} \\
{J1336+3414} & {8.4688(11)} & {1.507} & {71.97} & {77.98} & {1.67} & {--} & {--} & {\citet{2018ARep...62...63T}} \\
  J1336--20 & 19.3 & 0.184 & 316.82 & 41.10 & -- & -- & -- & \citet{GBNCC2015KA} \\
  J1354+2453 & 20.0(2) & 0.851 & 27.46 & 75.73 & -- & -- & -- & \citet{GBNCC2015KA} \\
{J1400+2125} & {11.214(3)} & {1.855} & {16.66} & {73.34} & {0.5} & {--} & {--} & {\citet{2018AA...618A..70T}} \\
  J1404+14 & 13.3(1.5) & -- & 0.73 & 68.97 & -- & -- & -- & \citet{2024AA...689A...1T} \\
  J1404--58 & 229(5) & -- & 312.45 & 3.52 & 1.1 & 0.22 & 4 & \citet{Keane2011} \\
  J1424--56 & 32.9(1.1) & 1.427 & 315.48 & 3.91 & 7 & 0.11 & 7 & \citet{2010MNRAS.401.1057K} \\
  J1429--64 & 151.6(5) & -- & 313.40 & --3.16 & -- & -- & -- & \citet{2025MNRAS.537.1070T} \\
  J1433+00 & 23.5 & -- & 349.75 & 53.79 & 2 & -- & -- & \citet{Deneva16} \\
  J1439+76 & 22.29(2) & 0.948 & 115.40 & 38.60 & 450 & -- & -- & \citet{GBNCC2015KA} \\
  J1444--6026 & 367.7(1.4) & 4.759 & 316.40 & --0.54 & 0.78 & 0.22 & 21 & \citet{McLaughlin2006} \\
  J1502+28 & 14(1.5) & 3.784 & 42.78 & 61.13 & -- & -- & -- & \citet{2018AA...618A..70T} \\
  J1513--5946 & 171.7(9) & 1.046 & 319.97 & --1.70 & 20 & 0.83 & 3.3 & \citet{2010MNRAS.401.1057K} \\
  J1525--2322 & 41.2(3) & 5.572 & 342.88 & 27.34 & -- & -- & -- & \citet{2025MNRAS.537.1070T} \\
  J1530+00 & 13.4(1.5) & -- & 5.10 & 43.61 & -- & -- & -- & \citet{2024AA...689A...1T} \\
  J1531--5557 & 56.6(6) & 2.920 & 323.99 & 0.24 & -- & -- & -- & \citet{2025MNRAS.537.1070T} \\
  J1533--5609 & 95.31(9) & 1.062 & 324.18 & --0.14 & -- & -- & -- & \citet{2025MNRAS.537.1070T} \\
  J1534--46 & 64.4(7.8) & 0.365 & 330.01 & 7.91 & -- & 0.14 & 25.5 & \citet{BS10} \\
  J1538+2345 & 14.909(1) & 3.449 & 37.32 & 52.39 & 129 & -- & -- & \citet{GBNCC2015KA} \\
  J1541+4703 & 19.4(7) & 0.278 & 75.50 & 51.39 & 8.1 & -- & -- & \citet{2023MNRAS.524.5132D} \\
  J1541--42 & 60(10) & -- & 333.49 & 10.23 & 7 & 0.15 & 4 & \citet{HTRU-RRATs} \\
  J1548--5229 & 366.0(5) & 4.850 & 328.14 & 1.48 & -- & -- & -- & \citet{2025MNRAS.544.1843T} \\
  J1549--57 & 17.7(3.5) & 0.738 & 325.13 & --2.35 & 73 & 0.21 & 4 & \citet{HTRU-RRATs} \\
  J1550+09 & 21(1.5) & -- & 19.30 & 44.35 & -- & -- & -- & \citet{2020BLPI...47..390L} \\
  J1554+18 & 24.0 & -- & 30.69 & 47.07 & 11 & -- & -- & \citet{Deneva16} \\
  J1554--5209 & 130.8(3) & 0.125 & 329.01 & 1.19 & 50.3 & 1.4 & 1.0 & \citet{2010MNRAS.401.1057K} \\
  J1555+01 & 18.5(1.5) & 0.577 & 10.45 & 38.73 & -- & -- & -- & \citet{2018ARep...62...63T} \\
  J1603+18 & 29.7 & 0.503 & 32.85 & 45.28 & 4 & -- & -- & \citet{Deneva16} \\
  J1605--07 & 4.8(1.0) & 1.810 & 3.54 & 31.65 & -- & -- & -- & \citet{2024AA...689A...1T} \\
  J1610--17 & 52.5(3.0) & -- & 355.61 & 24.11 & 13.6 & 0.23 & 5 & \citet{BS10} \\
  J1611--01 & 27.21(7) & 1.297 & 10.45 & 34.16 & 51 & -- & -- & \citet{GBNCC2015KA} \\
  \bottomrule
\end{tabular}
    \caption{-- {\it continued}}
\end{table*}
\addtocounter{table}{-1}
\begin{table*}
    \centering
    \small
\begin{tabular}{lllrrrrrl}
\toprule
      RRAT &   DM  &         $P$ & \multicolumn{1}{c}{$l$} & \multicolumn{1}{c}{$b$} & ${\cal B}$ & ${S}_{1400}$  & ${W}_{1400}$ & Reference \\
     &  (cm$^{-3}$~pc) &         (s) & \multicolumn{1}{c}{(\degr)} & \multicolumn{1}{c}{(\degr)} & (hr$^{-1}$) & (Jy) & (ms) & \\\midrule
  J1623--0841 & 59.79(2) & 0.503 & 5.77 & 27.37 & 35.77 & -- & -- & \citet{2013ApJ...763...80B} \\
  J1637--53 & 296(2) & -- & 332.78 & --4.19 & -- & -- & -- & \citet{2025MNRAS.544.1843T} \\
  J1641--51 & 250.4(7) & 5.514 & 334.95 & --3.19 & -- & -- & -- & \citet{2025MNRAS.537.1070T} \\
  J1647--3607 & 224(1) & 0.212 & 347.08 & 5.77 & 436.4 & -- & 9.9 & \citet{BS10} \\
  J1647--41 & 35.7(3) & -- & 343.08 & 2.49 & -- & -- & -- & \citet{2025MNRAS.544.1843T} \\
  J1648--51 & 201.4(2) & -- & 335.23 & --4.38 & -- & -- & -- & \citet{2025MNRAS.537.1070T} \\
  J1649--46 & 394(10) & -- & 339.65 & --0.76 & 0.3 & 0.135 & 16 & \citet{Keane2011} \\
  J1652--4406 & 786(10) & 7.707 & 341.56 & --0.09 & 0.7 & 0.04 & 64 & \citet{Keane2011} \\
  J1653--37 & 283(1) & -- & 346.48 & 3.92 & -- & -- & -- & \citet{2025MNRAS.544.1843T} \\
  J1654--2335 & 74.5(2.5) & 0.545 & 357.86 & 12.55 & 40.9 & 0.71 & 0.52 & \citet{Keane2011} \\
  J1655--40 & 92(1) & -- & 344.43 & 1.59 & -- & -- & -- & \citet{2025MNRAS.544.1843T} \\
  J1656+00 & 46.9 & 1.498 & 19.34 & 25.51 & -- & -- & 7.4 & \citet{Deneva16} \\
  J1703--38 & 375(12) & 6.443 & 347.41 & 2.04 & 3.2 & 0.16 & 9.0 & \citet{2010MNRAS.401.1057K} \\
  J1705--04 & 42.951(9) & 0.237 & 15.71 & 21.10 & 26 & -- & -- & \citet{GBNCC2015KA} \\
  J1707--4417 & 380(10) & 5.764 & 343.04 & --2.28 & 8.5 & 0.575 & 12.1 & \citet{2010MNRAS.401.1057K} \\
  J1709--43 & 228(20) & 0.897 & 343.57 & --2.36 & 7 & 0.24 & 3 & \citet{HTRU-RRATs} \\
  J1717+03 & 25.26 & 3.902 & 24.66 & 22.20 & 8 & -- & -- & \citet{Deneva16} \\
  J1720+00 & 46.0 & 3.357 & 22.74 & 20.43 & 33 & -- & -- & \citet{Deneva16} \\
  J1724--35 & 554.9(9.9) & 1.422 & 351.83 & --0.01 & 3.4 & 0.18 & 5.9 & \citet{2009MNRAS.395..410E} \\
  J1727--29 & 93(10) & -- & 356.97 & 2.80 & 0.9 & 0.16 & 7.2 & \citet{2010MNRAS.401.1057K} \\
  J1739--2521 & 186.4 & 1.818 & 2.33 & 3.03 & 22.64 & -- & -- & \citet{2017ApJ...840....5C} \\
  J1740+27 & 35.46(5) & 1.058 & 51.44 & 26.72 & -- & -- & -- & \citet{2018AA...618A..70T} \\
  J1748--3615 & 266.6(5) & 7.623 & 354.01 & --4.26 & -- & -- & -- & \citet{2025MNRAS.544.1843T} \\
  J1753--12 & 73.2(5.2) & 0.405 & 14.61 & 6.70 & 40.9 & 0.16 & 18.8 & \citet{BS10} \\
  J1753--38 & 168.4(1.3) & 0.667 & 352.28 & --6.37 & 26.3 & 0.44 & 4.8 & \citet{BS10} \\
  J1754--3014 & 89.7(7) & 1.320 & 359.86 & --2.33 & 0.6 & 0.16 & 16 & \citet{McLaughlin2006} \\
  J1807--11 & 152.4(4) & -- & 17.32 & 4.22 & -- & -- & -- & \citet{2025MNRAS.537.1070T} \\
  J1807--2557 & 385(10) & 2.764 & 4.99 & --2.65 & 6.2 & 0.41 & 4.0 & \citet{2010MNRAS.401.1057K} \\
  J1808--36 & 41.0(5) & -- & 355.39 & --8.15 & -- & -- & -- & \citet{2025MNRAS.537.1070T} \\
  J1816--2419 & 269.4(7) & 4.613 & 7.45 & --3.74 & -- & -- & -- & \citet{2025MNRAS.544.1843T} \\
  J1817--1932 & 214.5(2) & 1.229 & 11.72 & --1.59 & -- & -- & -- & \citet{2025MNRAS.544.1843T} \\
  J1819--1458 & 196.0(4) & 4.263 & 16.02 & 0.08 & 17.6 & 3.6 & 3 & \citet{McLaughlin2006} \\
  J1821--0031 & 111.1 & 4.441 & 28.98 & 6.53 & -- & -- & -- & \citet{2025RAA....25a4001H} \\
  J1825--33 & 43.2(2.0) & 1.271 & 0.31 & --9.70 & 14.4 & 0.36 & 16.5 & \citet{BS10} \\
  J1826--08 & 19.9(1.5) & -- & 22.37 & 1.67 & -- & -- & -- & \citet{2024AA...689A...1T} \\
  J1826--1419 & 160(1) & 0.771 & 17.40 & --1.14 & 1.06 & 0.52 & 2 & \citet{McLaughlin2006} \\
  J1828+0157 & 32.1 & 1.904 & 32.04 & 6.01 & -- & -- & -- & \citet{2025RAA....25a4001H} \\
  J1828--0003 & 193(3) & 3.807 & 30.29 & 5.03 & -- & -- & -- & \citet{2023RAA....23j4001Z} \\
  J1828--0038 & 70(2) & 2.426 & 29.72 & 4.87 & -- & -- & -- & \citet{2023RAA....23j4001Z} \\
  J1830+18 & 57.6(2.0) & -- & 47.10 & 12.76 & -- & -- & -- & \citet{2024AA...689A...1T} \\
  J1830--0231 & 150.7 & -- & 28.28 & 3.54 & -- & -- & -- & \citet{2025RAA....25a4001H} \\
  J1831--1141 & 46.1(2) & -- & 20.23 & --0.86 & -- & -- & -- & \citet{2025MNRAS.544.1843T} \\
  J1833+0050 & 190.9 & 0.904 & 31.68 & 4.27 & -- & -- & -- & \citet{2025RAA....25a4001H} \\
  J1836--0011 & 237.5 & 0.940 & 31.02 & 3.32 & -- & -- & -- & \citet{2025RAA....25a4001H} \\
  J1838+0414 & 154.2 & 1.331 & 35.22 & 4.83 & -- & -- & -- & \citet{eparent_thesis} \\
  J1838+50 & 21.81(1) & 2.577 & 79.82 & 22.74 & 3.9 & -- & -- & \citet{2021ApJ...922...43G} \\
  J1839--0141 & 293.2(6) & 0.933 & 30.01 & 1.96 & 0.61 & 0.1 & 15 & \citet{McLaughlin2006} \\
  J1840--0245 & 277(2) & 1.502 & 29.21 & 1.24 & -- & -- & -- & \citet{2023RAA....23j4001Z} \\
  J1840--0809 & 300.0(4) & 0.121 & 24.44 & --1.31 & -- & -- & -- & \citet{2025MNRAS.544.1843T} \\
  J1840--1419 & 19.4(1.4) & 6.598 & 18.94 & --4.12 & 46.0 & 1.7 & 2.6 & \citet{2010MNRAS.401.1057K} \\
  J1841+0328 & 153.1 & 0.445 & 34.86 & 3.85 & -- & -- & -- & \citet{eparent_thesis} \\
  J1841--0238 & 165.9 & 0.884 & 29.43 & 1.06 & -- & -- & -- & \citet{2025RAA....25a4001H} \\
  J1841--04 & 29(3) & -- & 27.49 & 0.09 & -- & -- & -- & \citet{2018AA...618A..70T} \\
  J1842+0114 & 307(8) & 4.140 & 32.98 & 2.61 & -- & -- & -- & \citet{2023RAA....23j4001Z} \\
  J1843+0527 & 261.1 & 2.035 & 36.91 & 4.19 & -- & -- & -- & \citet{eparent_thesis} \\
  J1843--0051 & 573(3) & 0.580 & 31.27 & 1.37 & -- & -- & -- & \citet{2023RAA....23j4001Z} \\
  J1843--0147 & 531.0 & -- & 30.44 & 0.95 & -- & -- & -- & \citet{2025RAA....25a4001H} \\
  J1843--0757 & 255(1) & 2.032 & 24.95 & --1.88 & -- & -- & 31.8 & \citet{2022MNRAS.512.1483B} \\
  J1845+0326 & 144(1) & 0.968 & 35.34 & 2.84 & -- & -- & -- & \citet{2023RAA....23j4001Z} \\
  J1845+0417 & 164(3) & 1.697 & 36.08 & 3.26 & -- & -- & -- & \citet{2023RAA....23j4001Z} \\
  J1845--0008 & 143(3) & 1.268 & 32.09 & 1.34 & -- & -- & -- & \citet{2023RAA....23j4001Z} \\
  \bottomrule
\end{tabular}
    \caption{-- {\it continued}}
\end{table*}
\addtocounter{table}{-1}
\begin{table*}
    \centering
    \small
\begin{tabular}{lllrrrrrl}
\toprule
      RRAT &   DM  &         $P$ & \multicolumn{1}{c}{$l$} & \multicolumn{1}{c}{$b$} & ${\cal B}$ & ${S}_{1400}$  & ${W}_{1400}$ & Reference \\
     &  (cm$^{-3}$~pc) &         (s) & \multicolumn{1}{c}{(\degr)} & \multicolumn{1}{c}{(\degr)} & (hr$^{-1}$) & (Jy) & (ms) & \\\midrule
  J1846--0257 & 237(7) & 4.477 & 29.71 & --0.20 & 1.1 & 0.2 & 15 & \citet{McLaughlin2006} \\
  J1847--0046 & 337(7) & -- & 31.83 & 0.46 & -- & -- & -- & \citet{2023RAA....23j4001Z} \\
  J1848+0009 & 393.4(4) & 4.708 & 32.79 & 0.62 & -- & -- & -- & \citet{2025MNRAS.544.1843T} \\
  J1848+1516 & 77.436(9) & 2.234 & 46.33 & 7.44 & -- & -- & -- & \citet{2018AA...618A..70T} \\
  J1848--1243 & 91.96(7) & 0.414 & 21.22 & --5.08 & 1.25 & 0.45 & 2 & \citet{McLaughlin2006} \\
  J1849+0619 & 110(1) & 2.011 & 38.35 & 3.29 & -- & -- & -- & \citet{2023RAA....23j4001Z} \\
  J1850+15 & 24.7(8.7) & 1.384 & 46.69 & 7.29 & -- & 0.2 & 31.8 & \citet{BS10} \\
  J1850--0004 & 154(1) & -- & 32.72 & 0.27 & -- & -- & -- & \citet{2023RAA....23j4001Z} \\
  J1851+0051 & 575(5) & 4.027 & 33.71 & 0.34 & -- & -- & -- & \citet{2023RAA....23j4001Z} \\
  J1853+0209 & 350(15) & -- & 35.04 & 0.61 & -- & -- & -- & \citet{2023RAA....23j4001Z} \\
  J1853+0353 & 379(2) & -- & 36.62 & 1.32 & -- & -- & -- & \citet{2023RAA....23j4001Z} \\
  J1853--0130 & 344(1) & 1.945 & 31.79 & --1.06 & -- & -- & -- & \citet{2023RAA....23j4001Z} \\
  J1854+0306 & 192.4(5.2) & 4.558 & 35.99 & 0.83 & 84 & 0.014 & -- & \citet{Keane2011} \\
  J1854--1557 & 150(17) & 3.453 & 19.02 & --7.95 & 25 & 0.05 & 65 & \citet{HTRU-RRATs} \\
  J1855+0033 & 554(1) & -- & 33.83 & --0.55 & -- & -- & -- & \citet{2023RAA....23j4001Z} \\
  J1855+0240 & 397(3) & 1.224 & 35.74 & 0.38 & -- & -- & -- & \citet{2023RAA....23j4001Z} \\
  J1855--0054 & 577(4) & -- & 32.58 & --1.28 & -- & -- & -- & \citet{2023RAA....23j4001Z} \\
  J1855--0154 & 417(1) & -- & 31.66 & --1.69 & -- & -- & -- & \citet{2023RAA....23j4001Z} \\
  J1855--0211 & 304(3) & -- & 31.45 & --1.91 & -- & -- & -- & \citet{2023RAA....23j4001Z} \\
  J1856+0029 & 234(3) & 0.376 & 33.98 & --0.98 & -- & -- & -- & \citet{2023RAA....23j4001Z} \\
  J1856+0528 & 307(2) & -- & 38.36 & 1.39 & -- & -- & -- & \citet{2023RAA....23j4001Z} \\
  J1857+0229 & 574(1) & 0.584 & 35.81 & --0.17 & -- & -- & -- & \citet{2023RAA....23j4001Z} \\
  J1857+0719 & 308.1 & 1.071 & 40.12 & 2.03 & -- & -- & -- & \citet{Patel2018} \\
  J1858+0453 & 429(1) & 3.761 & 38.12 & 0.59 & -- & -- & -- & \citet{2023RAA....23j4001Z} \\
  J1858--0113 & 280(4) & 1.532 & 32.70 & --2.21 & -- & -- & -- & \citet{2023RAA....23j4001Z} \\
  J1859+0239B & 624(4) & 0.849 & 36.19 & --0.54 & -- & -- & -- & \citet{2023RAA....23j4001Z} \\
  J1859+0251 & 286(3) & 3.580 & 36.40 & --0.51 & -- & -- & -- & \citet{2023RAA....23j4001Z} \\
  J1859+07 & 303.1(2.2) & -- & 40.02 & 1.51 & -- & 0.02 & 4.5 & \citet{Patel2018} \\
  J1859+0832 & 259(2) & -- & 41.44 & 2.11 & -- & -- & -- & \citet{2023RAA....23j4001Z} \\
  J1859--0233 & 164.2 & -- & 31.61 & --3.01 & -- & -- & -- & \citet{2025RAA....25a4001H} \\
  J1900+0732 & 226(1) & 1.709 & 40.64 & 1.48 & -- & -- & -- & \citet{2023RAA....23j4001Z} \\
  J1900+0908 & 264(4) & -- & 42.07 & 2.20 & -- & -- & -- & \citet{2023RAA....23j4001Z} \\
  J1900--0152 & 314(2) & 1.384 & 32.35 & --2.96 & -- & -- & -- & \citet{2023RAA....23j4001Z} \\
  J1902+0557 & 414(2) & -- & 39.54 & 0.17 & -- & -- & -- & \citet{2023RAA....23j4001Z} \\
  J1903+0319 & 307(3) & 1.854 & 37.23 & --1.11 & -- & -- & -- & \citet{2023RAA....23j4001Z} \\
  J1904+0100 & 146.3 & 1.309 & 35.27 & --2.37 & -- & -- & -- & \citet{2025RAA....25a4001H} \\
  J1904+0621 & 173(1) & 1.232 & 40.12 & --0.09 & -- & -- & -- & \citet{2023RAA....23j4001Z} \\
  J1905+0156 & 137(1) & 1.085 & 36.22 & --2.16 & -- & -- & -- & \citet{2023RAA....23j4001Z} \\
  J1905+0414 & 383.0 & 0.894 & 38.27 & --1.12 & -- & 0.036 & 3.3 & \citet{Patel2018} \\
  J1905+0558 & 472(1) & 0.846 & 39.79 & --0.30 & -- & -- & -- & \citet{2023RAA....23j4001Z} \\
  J1905+0849 & 257.8(2.3) & 1.034 & 42.32 & 1.01 & -- & -- & -- & \citet{2021RAA....21..107H} \\
  J1905+1200 & 183.5 & -- & 45.16 & 2.46 & -- & -- & -- & \citet{2025RAA....25a4001H} \\
  J1905--0128 & 100.3 & 1.071 & 33.23 & --3.81 & -- & -- & -- & \citet{2025RAA....25a4001H} \\
  J1906+0310 & 307.5 & -- & 37.45 & --1.86 & -- & -- & -- & \citet{2025RAA....25a4001H} \\
  J1906+0335 & 213(1) & 1.296 & 37.87 & --1.78 & -- & -- & -- & \citet{2023RAA....23j4001Z} \\
  J1907+0555 & 150(5) & 3.159 & 40.02 & --0.85 & -- & -- & -- & \citet{2023RAA....23j4001Z} \\
  J1908+0911 & 132(4) & 5.166 & 43.00 & 0.50 & -- & -- & -- & \citet{2023RAA....23j4001Z} \\
  J1908+1351 & 180.4 & 3.175 & 47.20 & 2.55 & -- & -- & -- & \citet{eparent_thesis} \\
  J1909+0310 & 110.7(4) & -- & 37.86 & --2.63 & -- & -- & -- & \citet{2025MNRAS.544.1843T} \\
  J1909+0641 & 36.7(2) & 0.742 & 40.94 & --0.94 & 67 & 0.082 & -- & \citet{2013ApJ...772...50N} \\
  J1910--0016 & 110.1 & 2.064 & 34.83 & --4.36 & -- & -- & -- & \citet{2025RAA....25a4001H} \\
  J1911+00 & 100(3) & 6.940 & 35.81 & --4.25 & 0.31 & 0.25 & 5 & \citet{McLaughlin2006} \\
  J1911+0310 & 167.7(8) & 1.333 & 38.02 & --2.97 & -- & -- & -- & \citet{2023RAA....23j4001Z} \\
  J1911+0906 & 24.3 & 16.926 & 43.32 & --0.30 & -- & -- & -- & \citet{2025RAA....25a4001H} \\
  J1911+1017 & 162(2) & 1.337 & 44.32 & 0.34 & -- & -- & -- & \citet{2023RAA....23j4001Z} \\
  J1911+1440 & 87.2 & 0.582 & 48.23 & 2.34 & -- & -- & -- & \citet{2025RAA....25a4001H} \\
  J1911+1525 & 299.8 & 3.282 & 48.93 & 2.62 & -- & -- & -- & \citet{2025RAA....25a4001H} \\
  J1911--2020 & 71.3(3) & 4.468 & 16.68 & --13.33 & -- & -- & -- & \citet{2025MNRAS.537.1070T} \\
  J1912+1000 & 147(4) & 3.053 & 44.25 & --0.13 & -- & -- & -- & \citet{2023RAA....23j4001Z} \\
  J1913+0400 & 125.4 & 0.391 & 39.07 & --3.18 & -- & -- & -- & \citet{2025RAA....25a4001H} \\
  J1913+1058 & 175.9 & -- & 45.14 & 0.25 & -- & -- & -- & \citet{2025RAA....25a4001H} \\
  \bottomrule
\end{tabular}
    \caption{-- {\it continued}}
\end{table*}
\addtocounter{table}{-1}
\begin{table*}
    \centering
    \small
\begin{tabular}{lllrrrrrl}
\toprule
      RRAT &   DM  &         $P$ & \multicolumn{1}{c}{$l$} & \multicolumn{1}{c}{$b$} & ${\cal B}$ & ${S}_{1400}$  & ${W}_{1400}$ & Reference \\
     &  (cm$^{-3}$~pc) &         (s) & \multicolumn{1}{c}{(\degr)} & \multicolumn{1}{c}{(\degr)} & (hr$^{-1}$) & (Jy) & (ms) & \\\midrule
  J1913+1330 & 175.64(6) & 0.923 & 47.42 & 1.38 & 4.7 & 0.46 & 2 & \citet{McLaughlin2006} \\
  J1914+0218 & 161.4 & 2.018 & 37.65 & --4.13 & -- & -- & -- & \citet{2025MNRAS.544.1843T} \\
  J1914+1053 & 108.0 & -- & 45.20 & --0.05 & -- & -- & -- & \citet{2025RAA....25a4001H} \\
  J1915+0639 & 212.32(5) & 0.644 & 41.65 & --2.37 & -- & -- & -- &  \citet{2022ApJ...924..135P} \\
  J1915+1045 & 123(3) & 1.546 & 45.23 & --0.38 & -- & -- & -- & \citet{2023RAA....23j4001Z} \\
  J1915--11 & 91.06(8) & 2.177 & 25.24 & --10.39 & 26 & -- & -- & \citet{GBNCC2015KA} \\
  J1916+0937 & 186(2) & 7.368 & 44.28 & --1.02 & -- & -- & -- & \citet{2023RAA....23j4001Z} \\
  J1916+1142A & 260(8) & -- & 46.24 & --0.26 & -- & -- & -- & \citet{2023RAA....23j4001Z} \\
  J1916+1142B & 318(8) & 1.188 & 46.24 & --0.26 & -- & -- & -- & \citet{2023RAA....23j4001Z} \\
  J1917+0834 & 101(3) & 2.933 & 43.47 & --1.74 & -- & -- & -- & \citet{2023RAA....23j4001Z} \\
  J1918+0342 & 174(5) & -- & 39.31 & --4.28 & -- & -- & -- & \citet{2023RAA....23j4001Z} \\
  J1918+0523 & 102.0 & 3.657 & 40.83 & --3.56 & -- & -- & -- & \citet{2025RAA....25a4001H} \\
  J1918+1514 & 134(2) & -- & 49.58 & 0.97 & -- & -- & -- & \citet{2023RAA....23j4001Z} \\
  J1918--0449 & 116.1(4) & 2.479 & 31.70 & --8.24 & 54.5 & -- & -- & \citet{2022ApJ...934...24C} \\
  J1919+1113 & 288(2) & 0.766 & 46.09 & --1.02 & -- & -- & -- & \citet{2023RAA....23j4001Z} \\
  J1919+1745 & 142.3(2) & 2.081 & 51.90 & 1.99 & 320 & 0.012 & -- & \citet{2013ApJ...772...50N} \\
  J1921+0851 & 101(2) & 0.957 & 44.20 & --2.50 & -- & -- & -- & \citet{2023RAA....23j4001Z} \\
  J1921+1006 & 362(8) & 3.345 & 45.37 & --2.04 & -- & -- & -- & \citet{2023RAA....23j4001Z} \\
  J1921+1227 & 259(2) & 1.598 & 47.40 & --0.85 & -- & -- & -- & \citet{2023RAA....23j4001Z} \\
  J1921+1629 & 105(2) & -- & 51.01 & 0.96 & -- & -- & -- & \citet{2023RAA....23j4001Z} \\
  J1921+1632 & 164(2) & 0.493 & 51.03 & 1.02 & -- & -- & -- & \citet{2023RAA....23j4001Z} \\
  J1924+1006 & 178.1 & 4.620 & 45.69 & --2.63 & -- & -- & -- & \citet{eparent_thesis} \\
  J1924+1446 & 336(3) & 1.090 & 49.85 & --0.51 & -- & -- & -- & \citet{2023RAA....23j4001Z} \\
  J1924+1734 & 49(3) & -- & 52.32 & 0.80 & -- & -- & -- & \citet{2023RAA....23j4001Z} \\
  J1925--16 & 88(20) & 3.886 & 22.13 & --14.54 & 6.5 & 0.16 & 10 & \citet{HTRU-RRATs} \\
  J1927+1126 & 55.2 & 5.889 & 47.23 & --2.69 & -- & -- & -- & \citet{2025RAA....25a4001H} \\
  J1927+1849 & 200(3) & 0.312 & 53.74 & 0.81 & -- & -- & -- & \citet{2023RAA....23j4001Z} \\
  J1927+1940 & 347(2) & -- & 54.43 & 1.31 & -- & -- & -- & \citet{2023RAA....23j4001Z} \\
  J1928+15 & 242 & 0.403 & 50.64 & --1.03 & 4 & 0.18 & 5 & \citet{Deneva09} \\
  J1928+17 & 136.0(1) & 0.290 & 52.64 & --0.09 & 78 & -- & 1.1 & \citet{2022ApJ...924..135P} \\
  J1929+1155 & 81.2 & 3.217 & 47.85 & --2.80 & -- & -- & -- & \citet{eparent_thesis} \\
  J1930+1713 & 488.9 & -- & 52.69 & --0.61 & -- & -- & -- & \citet{2025RAA....25a4001H} \\
  J1930--1856 & 63.143(9) & 1.761 & 19.92 & --16.95 & -- & -- & -- & \citet{2025MNRAS.537.1070T} \\
{J1931+4229} & {50.987(6)} & {3.921} & {75.15} & {11.23} & {8} & {--} & {--} & {\citet{2021ApJ...922...43G}} \\
  J1932+2126 & 126(3) & -- & 56.61 & 1.02 & -- & -- & -- & \citet{2023RAA....23j4001Z} \\
  J1933+2315 & 216.4 & 1.167 & 58.29 & 1.73 & -- & -- & -- & \citet{2025RAA....25a4001H} \\
  J1933+2401 & 185(3) & -- & 58.95 & 2.11 & -- & -- & -- & \citet{2023RAA....23j4001Z} \\
  J1934+2341 & 252(2) & -- & 58.71 & 1.86 & -- & -- & -- & \citet{2023RAA....23j4001Z} \\
  J1935+1841 & 290(3) & 5.529 & 54.45 & --0.77 & -- & -- & -- & \citet{2023RAA....23j4001Z} \\
  J1935+1901 & 365(2) & 0.897 & 54.83 & --0.77 & -- & -- & -- & \citet{2023RAA....23j4001Z} \\
  J1938+1748 & 56(1) & 7.106 & 54.08 & --1.91 & -- & -- & -- & \citet{2023RAA....23j4001Z} \\
  J1940+2203 & 59(9) & 11.906 & 58.05 & --0.30 & -- & -- & -- & \citet{2023RAA....23j4001Z} \\
  J1940+2231 & 198(7) & 5.682 & 58.47 & --0.09 & -- & -- & -- & \citet{2023RAA....23j4001Z} \\
  J1942+2604 & 161.0 & 2.642 & 61.74 & 1.35 & -- & -- & -- & \citet{2025RAA....25a4001H} \\
  J1943+09 & 46(2) & -- & 47.59 & --7.01 & -- & -- & -- & \citet{2024AA...689A...1T} \\
  J1944--10 & 31.01(3) & 0.409 & 29.53 & --16.29 & 180 & -- & -- & \citet{GBNCC2015KA} \\
  J1945+2357 & 87.5 & 4.718 & 60.27 & --0.35 & 54 & 0.101 & 4 & \citet{Deneva09} \\
  J1948+2314 & 184(3) & 1.471 & 59.98 & --1.27 & -- & -- & -- & \citet{2023RAA....23j4001Z} \\
  J1948+2438 & 450(4) & 1.903 & 61.16 & --0.51 & -- & -- & -- & \citet{2023RAA....23j4001Z} \\
  J1951+2329 & 260.0 & 1.826 & 60.47 & --1.61 & -- & -- & -- & \citet{2025RAA....25a4001H} \\
  J1952+30 & 188.8(6) & 1.666 & 66.52 & 1.65 & -- & 0.033 & 5.7 & \citet{Patel2018} \\
  J1953--6112 & 43.0(1) & 0.461 & 335.70 & --30.90 & -- & -- & -- & \citet{2025MNRAS.544.1843T} \\
  J1956+2911 & 265(2) & 3.816 & 66.00 & 0.25 & -- & -- & -- & \citet{2023RAA....23j4001Z} \\
  J1956+3544 & 153.5(1) & 0.876 & 71.60 & 3.68 & -- & -- & -- & \citet{2025MNRAS.544.1843T} \\
  J1956--28 & 45.69(1) & 0.260 & 13.22 & --25.59 & 120 & -- & -- & \citet{GBNCC2015KA} \\
  J2001+4209 & 153(2) & -- & 77.62 & 6.14 & -- & -- & -- & \citet{2023RAA....23j4001Z} \\
  J2005+3154 & 225(1) & -- & 69.30 & 0.09 & -- & -- & -- & \citet{2023RAA....23j4001Z} \\
  J2005+3156 & 337(2) & 2.146 & 69.35 & 0.08 & -- & -- & -- & \citet{2023RAA....23j4001Z} \\
  J2007+13 & 67.4(2.0) & -- & 53.41 & --10.28 & -- & -- & -- & \citet{2024AA...689A...1T} \\
  J2007+20 & 67.0(4) & 4.634 & 59.72 & --6.41 & 77 & -- & -- & \citet{GBNCC2015KA} \\
  J2008+3758 & 143(1) & 4.352 & 74.71 & 2.90 & 2.4 & -- & -- & \citet{2023MNRAS.524.5132D} \\
  \bottomrule
\end{tabular}
    \caption{-- {\it continued}}
\end{table*}
\addtocounter{table}{-1}
\begin{table*}
    \centering
    \small
\begin{tabular}{lllrrrrrl}
\toprule
      RRAT &   DM  &         $P$ & \multicolumn{1}{c}{$l$} & \multicolumn{1}{c}{$b$} & ${\cal B}$ & ${S}_{1400}$  & ${W}_{1400}$ & Reference \\
     &  (cm$^{-3}$~pc) &         (s) & \multicolumn{1}{c}{(\degr)} & \multicolumn{1}{c}{(\degr)} & (hr$^{-1}$) & (Jy) & (ms) & \\\midrule
  J2014+3326 & 333(2) & 0.977 & 71.63 & --0.68 & -- & -- & -- & \citet{2023RAA....23j4001Z} \\
  J2019--07 & 24.7(1.5) & -- & 36.03 & --23.13 & -- & -- & -- & \citet{2024AA...689A...1T} \\
  J2030+3833 & 417(6) & -- & 77.69 & --0.43 & -- & -- & -- & \citet{2023RAA....23j4001Z} \\
  J2033+0042 & 37.8(1) & 5.013 & 45.88 & --22.20 & -- & 0.14 & 95.2 & \citet{BS10} \\
  J2044+3843 & 230.0 & -- & 79.47 & --2.50 & -- & -- & -- & \citet{2025RAA....25a4001H} \\
  J2047+12 & 36(2) & 2.925 & 58.96 & --18.60 & -- & -- & -- & \citet{2020BLPI...47..390L} \\
  J2051+1248 & 43.45(1) & 0.553 & 59.36 & --19.45 & -- & -- & -- & \citet{2018AA...618A..70T} \\
{J2105+19} & {34.47(3)} & {3.530} & {67.00} & {--18.18} & {--} & {--} & {--} & {\citet{2018AA...618A..70T}} \\
  J2105+6223 & 50.75(8) & 2.305 & 99.79 & 10.18 & 30 & -- & -- & \citet{GBNCC2015KA} \\
  J2113+73 & 42.4 & -- & 109.02 & 16.98 & 0.4 & -- & -- & \citet{2023MNRAS.524.5132D} \\
  J2119+40 & 72.6(3.0) & -- & 85.50 & --6.12 & -- & -- & -- & \citet{2024AA...689A...1T} \\
  J2129+4106 & 73.5 & 3.261 & 87.01 & --7.25 & -- & -- & -- & \citet{2025RAA....25a4001H} \\
{J2138+69} & {46.530(3)} & {0.220} & {107.60} & {12.95} & {0.3} & {--} & {--} & {\citet{2023MNRAS.524.5132D}} \\
{J2202+21} & {17.7473(19)} & {--} & {78.79} & {--26.25} & {--} & {--} & {--} & {\citet{2018AA...618A..70T}} \\
  J2209+22 & 46.3(8) & 1.777 & 79.90 & --27.79 & -- & -- & -- & \citet{2018AA...618A..70T} \\
{J2215+4524} & {18.5917(16)} & {2.723} & {96.39} & {--9.29} & {6.0} & {--} & {--} & {\citet{2023MNRAS.524.5132D}} \\
  J2218+2902 & 55.8(4) & 17.495 & 86.93 & --22.92 & -- & -- & -- & \citet{2025MNRAS.537.1070T} \\
  J2218--1229 & 26.8(6) & 0.163 & 47.56 & --51.41 & -- & -- & -- & \citet{2025MNRAS.544.1843T} \\
  J2221+81 & 39 & -- & 117.43 & 20.32 & 0.4 & -- & -- & \citet{2023MNRAS.524.5132D} \\
  J2225+35 & 51.8 & 0.942 & 92.11 & --18.45 & -- & -- & -- & \citet{2009ARep...53..561S} \\
  J2237+2828 & 38.1(4) & 1.077 & 90.34 & --25.79 & 1.0 & -- & -- & \citet{2023MNRAS.524.5132D} \\
  J2251+14 & 10.2(1.5) & -- & 83.81 & --39.56 & -- & -- & -- & \citet{2024AA...689A...1T} \\
  J2310+6706 & 97.7(1) & 1.945 & 113.35 & 6.14 & 60 & -- & -- & \citet{GBNCC2015KA} \\
  J2312+6931 & 71.6(1) & 0.813 & 114.44 & 8.29 & 60 & -- & -- & \citet{2018ApJ...859...93L} \\
  J2316+75 & 53.4 & -- & 116.98 & 13.99 & 0.5 & -- & -- & \citet{2023MNRAS.524.5132D} \\
  J2317--4746 & 15.9(3) & 1.733 & 338.25 & --62.39 & -- & -- & -- & \citet{2025MNRAS.544.1843T} \\
{J2325--0530} & {14.9580(8)} & {0.869} & {75.58} & {--60.20} & {103} & {--} & {--} & {\citet{GBNCC2015KA}} \\
  J2337--04 & 15.3(1.5) & -- & 81.36 & --61.38 & -- & -- & -- & \citet{2024AA...689A...1T} \\
{J2355+1523} & {26.924(16)} & {1.094} & {103.66} & {--45.39} & {6.0} & {--} & {--} & {\citet{2023MNRAS.524.5132D}} \\
  J2359+06 & 19.8(1.5) & -- & 100.43 & --54.04 & -- & -- & -- & \citet{2024AA...689A...1T} \\
  \bottomrule
\end{tabular}
    \caption{-- {\it continued}}
\end{table*}


\begin{table*}
    \centering
    \small
\begin{tabular}{llllllrrrr}
\toprule
      RRAT &  R.A. (J2000) & Decl. (J2000) &           $P$ & $\dot{P}$ & Epoch & $\log B_{\mathrm{s}}$ & $\log B_{\mathrm{LC}}$ & $\log \dot{E}$ & $\log \tau$  \\
      &   (h m s) &  (\degr \, \arcmin \, \arcsec) &           (s) &  ($10^{-15}$) & (MJD) &  &   &  &  \\
\midrule
{J0054+6650} & {00:54:55.412(14)} & {+66:50:23.81(8)} & {1.390218110074(8)} & {5.5532(8)} & {59478} & {12.4} & {0.8} & {31.9} & {6.6} \\
{J0102+5356} & {01:02:57.786(12)} & {+53:56:11.86(15)} & {0.354299198996(4)} & {0.5203(5)} & {59545} & {11.6} & {2.1} & {32.7} & {7.0} \\
J0139+3336  &  01:39:57.23(4) &  +33:36:59.7(9) &  1.2479609557(1) &   2.064(8) &  57901 & 12.2 & 0.9 & 31.6 & 7.0 \\
J0201+7005 &  02:01:41.344(7) & +70:05:18.11(6) & 1.349184471847(9) &   5.514(1) &  56777 & 12.4 & 1.0 & 31.9 & 6.6 \\
J0302+2252 &  03:02:31.990(4) &  +22:52:12.1(2) & 1.207164839778(2) &  0.0825(1) &  57811 & 11.5 & 0.2 & 30.3 & 8.4 \\
J0402--6542 & 04:02:52.27(3) & --65:42:43.41(16) & 3.03352298461(7) & 5.601(2) & 59581 & 12.6 & 0.1 & 30.9 & 6.9 \\
J0628+0909 &  06:28:36.183(5) &  +09:09:13.9(3) & 1.241421391299(3) &  0.5479(2) &  54990 & 11.9 & 0.6 & 31.0 & 7.6 \\
J0736--6304 &  07:36:20.01(27) &   --63:04:16(2) &   4.8628739612(7) &   151.9(2) &  56212 & 13.4 & 0.3 & 31.7 & 5.7 \\
{J0746+5514} & {07:46:47.4(2)} & {+55:14:37(2)} & {2.8936675025(4)} & {10.19(5)} & {59558} & {12.7} & {0.1} & {31.2} & {6.7} \\
J0847--4316 &   08:47:57.33(5) &  --43:16:56.8(7) &    5.977492737(7) &  119.94(2) &  53816 & 13.4 & 0.1 & 31.3 & 5.9 \\
J0912--3851 &   09:12:42.70(2) &    --38:51:03(1) &    1.526085076(3) &    3.59(5) &  55093 & 12.4 & 0.8 & 31.6 & 6.8 \\
{J1006+3015} & {10:06:34.44(7)} & {+30:15:46(2)} & {3.0663651860(2)} & {5.52(2)} & {59476} & {12.6} & {--0.1} & {30.9} & {6.9} \\
J1048--5838 &  10:48:12.56(1) & --58:38:19.02(10) &  1.23130477663(4) & 12.19375(7) &  53510 & 12.6 & 1.3 & 32.4 & 6.2 \\
J1108--5946 & 11:07:58.56(23) & --59:47:01.1(12) & 1.516531549(3) & $<$0.4 & 60001 & $<$11.9 & $<$0.8 & $<$30.7 & $>$7.8 \\
J1130+0921 & 11:30:55.0(5) & +09:21:09(14) & 4.796636974(6) & 2.9(5) & 59180 & 12.6 & --0.5 & 30.0 & 7.4 \\
J1226--3223 &  12:26:46.63(4) &    --32:23:01(1) &   6.1930040852(5) &    7.05(1) &  56114 & 12.8 & --0.5 & 30.0 & 7.1 \\
J1317--5759 &   13:17:46.29(3) &  --57:59:30.5(3) &   2.6421985132(5) &   12.56(3) &  53911 & 12.8 & 0.5 & 31.4 & 6.5 \\
J1319--4536 & 13:19:48.31(6) & --45:36:03.0(8) & 1.8709058202(2) & 6.975(3) & 59369 & 12.6 & 0.7 & 31.6 & 6.6 \\
{J1336+3414} & {13:36:33.953(18)} & {+34:14:37.8(2)} & {1.50660326853(3)} & {0.111(3)} & {59527} & {11.6} & {--0.1} & {30.1} & {8.3} \\
J1354+2453 & 13:54:13.383(3) & +24:53:46.16(6) & 0.851063907874(6) & 0.14007(7) & 58521 & 11.5 & 0.7 & 30.9 & 8.0 \\
{J1400+2125} & {14:00:14.19(3)} & {+21:25:40.0(5)} & {1.85546492484(6)} & {1.358(7)} & {59481} & {12.2} & {0.1} & {30.9} & {7.3} \\
J1444--6026 &   14:44:06.02(7) &  --60:26:09.4(4) &   4.7585755679(2) &   18.542(8) &  53893 & 13.0 & --0.1 & 30.8 & 6.6 \\
J1513--5946 &   15:13:44.78(1) &  --59:46:31.9(7) & 1.046117156733(8) &   8.5284(4) &  54909 & 12.5 & 1.4 & 32.5 & 6.3 \\
J1538+2345 &   15:38:06.07(2) &   +23:45:04.0(2) &  3.44938495332(9) &    6.89(1) &  56745 & 12.7 & 0.0 & 30.8 & 6.9 \\
J1541+4703 & 15:41:05.54(2) & +47:03:03.7(3) & 0.2777006928933(3) & 0.2102(9) & 59211 & 11.3 & 2.0 & 32.6 & 7.3 \\
J1554--5209 &   15:54:27.15(2) &  --52:09:39.3(4) & 0.1252295584025(7) & 2.29442(5) &  55039 & 11.7 & 3.4 & 34.7 & 5.9 \\
J1623--0841 & 16:23:42.69(1) &  --08:41:36.6(5) &   0.5030150056(1) &   1.9556(7) &  55079 & 12.0 & 1.9 & 32.8 & 6.6 \\
J1647--3607 &   16:47:46.51(2) &    --36:07:04(1) &  0.21231640921(5) &   0.129(2) &  54984 & 11.3 & 2.2 & 32.7 & 7.4 \\
J1652--4406 &    16:52:59.5(2) &    --44:06:05(4) &    7.707183007(4) &     9.5(2) &  54947 & 12.9 & --0.7 & 29.9 & 7.1 \\
J1707--4417 &   17:07:41.41(3) &    --44:17:19(1) &    5.763777003(4) &   11.65(2) &  54999 & 12.9 & --0.4 & 30.3 & 6.9 \\
J1709--43 & 17:09:47(39) & --43:54(7) & 0.8968609868 & 24.162 & 56800 & 12.7 & 1.8 & 33.1 & 5.8 \\
J1739--2521 &  17:39:32.63(5) &   --25:21:56(15) &   1.8184611929(2) &    0.24(2) &  55631 & 11.8 & 0.0 & 30.2 & 8.1 \\
J1754--3014 &  17:54:30.18(4) &    --30:15:03(5) &   1.3204904144(3) &    4.43(2) &  55292 & 12.4 & 1.0 & 31.9 & 6.7 \\
J1807--2557 &   18:07:13.66(1) &    --25:57:20(5) &  2.76419486975(4) &   4.994(2) &  54984 & 12.6 & 0.2 & 31.0 & 6.9 \\
J1817--1932 & 18:17:12.6(1) & --19:32:48(2) & 1.22912(2) & $<$6 & 59769 & $<$12.4 & $<$1.1 & $<$32.1 & $>$6.5 \\
J1819--1458 &   18:19:34.16(1) & --14:58:03.57(1) &   4.2632901504(1) & 562.717(4) & 55996 & 13.7 & 0.8 & 32.5 & 5.1 \\
J1826--1419 &  18:26:42.391(4) &  --14:19:21.6(3) & 0.770620171033(7) &   8.7841(2) &  54053 & 12.4 & 1.7 & 32.9 & 6.1 \\
J1839--0141 & 18:39:06.985(9) &  --01:41:56.0(2) &  0.93326558076(2) &   5.944(1) &  55467 & 12.4 & 1.4 & 32.5 & 6.4 \\
J1840--1419 &   18:40:33.04(1) &  --14:19:06.5(9) &   6.5975625223(1) &   6.353(1) & 55074 & 12.8 & --0.7 & 30.0 & 7.2 \\
J1843--0757 & 18:43:33.06(2) & --07:57:33(2) & 2.03194008516(9) & 4.13(3) & 58743 & 12.5 & 0.5 & 31.3 & 6.9 \\
J1846--0257 &   18:46:15.49(4) & --02:57:36.0(1.8) &   4.4767225398(1) & 160.587(3) &  53039 & 13.4 & 0.4 & 31.9 & 5.6 \\
J1848+1516 &   18:48:56.13(2) &   +15:16:44.1(4) &  2.23376977466(5) &   1.6813(8) &  57655 & 12.3 & 0.2 & 30.8 & 7.3 \\
J1848--1243 &  18:48:18.03(1) &    --12:43:30(1) &   0.4143833544(2) &   0.4405(8) &  55595 & 11.6 & 1.7 & 31.4 & 7.2 \\
J1854+0306 &   18:54:02.98(3) &    +03:06:14(1)  &   4.5578200962(1) & 145.125(6) &  54944 & 13.4 & 0.4 & 31.8 & 5.7 \\
J1854--1557 &    18:54:53.6(1) &   --15:57:47(14) &   3.4531211813(7) &    4.52(4) &  55124 & 12.6 & 0.0 & 30.6 & 7.1 \\
J1909+0641 &  19:09:29.052(4) &   +06:41:25.8(2) & 0.741761952452(6) &   3.2239(7) &  54870 & 12.2 & 1.5 & 32.5 & 6.6 \\
J1911--2020 & 19:11:16.05(8) & --20:20:02(9) & 4.4679211203(2) & 6.726(8) & 60098 & 12.7 & --0.2 & 30.5 & 7.0 \\
J1913+1330 &   19:13:17.97(1) &  +13:30:32.78(4) &  0.92339138665(2) &   8.6776(2) & 55090 & 12.5 & 1.5 & 32.6 & 6.2 \\
J1915+0639 & 19:15:54.327(2) & +06:39:46.21(4) & 0.64414015325(3) & 1.8435(4) & 57374 & 12.0 & 1.6 & 32.4 & 6.7 \\
J1919+1745 &  19:19:43.342(4) &  +17:45:03.79(8) & 2.081343459724(9) &    1.705(4) &  55320 & 12.3 & 0.3 & 30.9 & 7.3 \\
J1930--1856 & 19:30:41.88(9) & --18:56:28.5(12) & 1.76083292621(3) & 0.593(7) & 59581 & 12.0 & 0.2 & 30.6 & 7.7 \\
{J1931+4229} & {19:31:10.87(7)} & {+42:29:16.1(10)} & {3.9210375131(2)} & {31.14(5)} & {59520} & {13.0} & {0.0} & {31.3} & {6.3} \\
J2033+0042 &   20:33:31.12(2) &   +00:42:24.1(9) &  5.01340011141(8) &   9.693(2) &  57600 & 12.9 & --0.3 & 30.5 & 6.9 \\
J2051+1248 & 20:51:29.66(2) & +12:48:21.5(6) & 0.55316745256(2) & $<$0.025 & 57811 & $<$11.1 & $<$0.8 & $<$30.8 & $>$8.5 \\
J2105+6223 &   21:05:12.93(2) &   +62:23:05.5(1) &  2.30487883766(4) &   5.219(6) &  56774 & 12.5 & 0.4 & 31.2 & 6.8 \\
{J2215+4524} & {22:15:46.847(13)} & {+45:24:43.3(2)} & {2.72313605772(5)} & {5.223(5)} & {59534} & {12.6} & {0.2} & {31.0} & {6.9} \\
J2237+2828 & 22:37:29.41(4) & +28:28:40(4) & 1.0773950914(7) & $<$1.2 & 59289 & $<$12.1 & $<$1.0 & $<$31.5 & $>$7.2 \\
J2310+6706 &    23:10:42.0(3) &  +67:06:52.1(10) &    1.944788973(1) &   0.076(4) &  57225 & 11.6 & --0.3 & 29.6 & 8.6 \\
J2312+6931 & 23:12:38.93(5) & +69:31:04.0(3) & 0.81337477832(2) & 0.63(1) & 56500 & 11.8 & 1.1 & 31.7 & 7.3 \\
J2325--0530 &    23:25:15.3(1) &    --05:30:39(4) & 0.868735115026(9) &   1.029(2) &  56774 & 12.0 & 1.1 & 31.8 & 7.1 \\
J2355+1523 & 23:55:48.62(8) & +15:23:19(2) & 1.09439626467(5) & 0.41(2) & 59121 & 11.8 & 0.7 & 31.1 & 7.6 \\
\bottomrule
\end{tabular}

    \caption{Observed and derived parameters for currently known RRATs with measured values of $\dot{P}$. For each RRAT, we list its timing-derived position spin period, period derivative and timing epoch. Figures in parentheses are 1$\sigma$ uncertainties in the least significant digit of each of the fitted parameters. The derived parameters are  surface magnetic field, $B_{\rm s}$ (G), magnetic field at the light cylinder, $B_{\rm LC}$ (G), spin-down luminosity, $\dot{E}$ (erg~s$^{-1}$) and characteristic age, $\tau$ (yr). For sources where $\dot{P}$ is not detected at a significance $> 3\sigma$, we quote upper limits for $B_{\mathrm{s}}$, $B_{\mathrm{LC}}$, and $\dot{E}$ (and a lower limit for $\tau$) based on the $1\sigma$ threshold $\dot{P} + \sigma_{\dot{P}}$.}
    \label{tab:RRATspindown}
\end{table*}

\bsp	
\label{lastpage}
\end{document}